\documentclass[aps,prb,preprint,showpacs]{revtex4}
\usepackage{natbib}
\usepackage{epsfig}
\usepackage{color}

\begin{document}

\title{
Spin-1/2 operators exactly mapped to spinful canonical Fermi
  operators present in two bands.
}
\author{Zsolt~Gulacsi} 
\affiliation{Department of Theoretical Physics, University of
Debrecen, H-4010 Debrecen, Bem ter 18/B, Hungary}
\date{\today }

\begin{abstract}
Recently it has been shown that the quantum spin-1/2 spin operators can
be exactly transformed not only in spinless, but also in spinful canonical
Fermi operators in 1D [\cite{JW1}], and 2D [\cite{JW2}] as well. In this
paper, using the same technique based on an extended Jordan-Wigner
transformation, we show in 1D, that the quantum spin-1/2 operators can be
exactly transformed also in spinful canonical Fermi operators of spin-1/2
fermions that are belong to two bands.
\end{abstract}

\maketitle


\section{Introduction}

Mapping spin models to fermionic models transforms a system
description to
another type of description which analyzes the same model, but from a completely
different point of view using a completely different language. This change is
able to emphasize aspects that have not been observed previously, or underlines
new properties that help considerably the understanding of the physical
behavior and its mathematical description. Hence a such type of transformation
equally can be applied for solving models, or e.g. for information transfer
from one side of the mapping to the another side.

On this line, the
Jordan-Wigner transformation \cite{JW3}, as is usually considered since
1928, maps the quantum spin-1/2 operators to spinless Fermi operators in 1D.
It represents a pillar of theoretical physics for almost one century, see
e.g. Ref.[\cite{JW4}].
After sixty years of its appearance, maintaining the quantum spin-1/2 to
spinless fermion operator interconnection, it has been generalized to two
dimensions \cite{JW5,JW6,JW7}, and at the beginning of the second millennium
also extended to $S > 1/2$ case \cite{JW8}. After again one quarter of century
it was shown both in one and two dimensions that the quantum spin-1/2
operators can be exactly mapped not only in spinless fermions, but also in
spinful canonical Fermi operators \cite{JW1,JW2} opening the doors of the
transformation (mapping) possibility of spin models (systems) constructed from
quantum spin-1/2 operators to realistic fermionic systems (note that spinless
fermion is in fact a mathematical abstraction).

Up to this moment the transformation of quantum spin-1/2 operators to fermionic
operators has been performed for fermions belonging to one single band. In the
presented paper we show that the same quantum spin-1/2 operators can be as well
exactly transformed in spinful canonical Fermi operators describing spin-1/2
fermions present in two bands. The used technique is similar to that applied in
Refs.[\cite{JW1,JW2}], and is based on further extensions of the Jordan-Wigner
transformation. The presence of a such kind of transformation gives the
possibility to map quantum spin-1/2 models (systems) to realistic two
band fermionic systems as well.

Concerning the importance and the motivations of the presented transformation,
I mention the following aspects:
i) First the main information provided is that it can be
done, hence a such kind of exact transformation exists. When we transform spin
operators, in spinless fermion operators we show, that the spin operators are
able to cover by mapping global fermionic characteristics, i.e. the
anticommutation relations, which determines the fermion particle statistics.
This provides the impression that spin operators can done only this, and are
unable to touch fermionic characteristics that manifestly influence local
properties via spin. So at this level is seems that spin operators do not
see intrinsic (fermionic) spin degrees of freedom, hence loose to feel the
complexity of the spin dynamics at the fermionic side. When we show that
the spin operators can be transformed in spinful
canonical Fermi operators (building one band), we prove that this is not true.
The spin operators see and feel also this local characteristic. If we stop at
this stage we could conclude, that this is the maximum that the spin operators
are able to provide at the level local characteristics. On this
background, if we introduce two spinful fermions, we increase considerably the
fermionic degrees of freedom that influence local characteristics (e.g. we have
now 4 local occupation possibilities instead of 2, we capture new symmetries
\cite{JW3a}, etc.). Now if we
demonstrate that exact mapping is also possible in this case in between the
spin operators and this two bands spinful canonical Fermi operators, we show
that the spin operators are able to feel much more local information that we
conjectured, and have seen before. ii) Spinless fermion is in fact a
mathematical abstraction, and as such, a) provides the mathematical tool to
solve a spin model exactly (by the exact transformation of a spin model to a
simpler mathematical model whose exact solution is trivially seen, or it is
known), or b) allows the preliminary, or starting point numerical study of a
fermionic system without to encounter the troubling sign problem
\cite{JW3aa}. This stage do not allows to map spin systems to realistic
fermionic systems.

Mappings containing spinful fermions, even if not contains integrable
system on both sides, are extremely
useful, since allow the known information transfer from one side of the mapping
to the another side, increasing the knowledge accumulation of a studied
problem. For this to be possible, spinful fermions are needed, because all
realistic fermions hold spin. This level is attained by the transformation
described in Refs.[1,2], which also couples in the mapping fermionic systems
connected to explicit spin characteristics (as e.g. many-body spin-orbit
coupling, spin dependent hoppings \cite{JW30}  or interactions containing
even spin-flip possibilities \cite{JW300},
spin dependent pairings \cite{JW3b}, systems able to provide spin liquids
\cite{JW3c}). If one remains at this
stage, one could argue, that in fact realistic systems are of multiband type,
so the coopted one-band fermionic system is in fact not realistic. On this
line, one observes that during a description involving several bands, the
multiband system usually is projected in a minimal two band system, at the
level of which the study is done. This shows that the spin operator mapping to
two band spinful fermionic operators opens supplementary doors in coopting
realistic fermionic systems in this mapping (as potential possibilities I
mention besides the two band versions of the one band spinful fermion
opportunities mentioned above, e.g. multiband systems able to
provide spin dependent topological phases \cite{JW3d,JW3e,JW3f}).
iii) There is present another aspect related to integrability, which is
also intensively analyzed nowadays \cite{JW3g}. E.g. if the spin to two
bands exact transformation has an integrable system on the spin side, it
provides a two bands exactly solvable system on the fermionic side, which is
extremely rare. The transformation described in this paper, 
is applied for exemplification to the 1D Heisenberg model on the spin side,
hence provides a multitude of examples in this direction. iv) Given mainly by
the richer local properties of the two band systems, there are strongly spin
dependent two band models/systems whose behavior cannot be described by
one band models (e.g. periodic Anderson model \cite{JW3h}). According to the
presented transformation, also such type of behaviors could be coopted by spin
models containing one spin-1/2 spin per site. So increasing the local diversity
on the fermionic side to 4 particles per site at the level of local occupancy
possibilities, this is also supported by one spin-1/2 spin per site on the
spin side of the mapping, which is a new, and quite interesting property.
v.) In my knowledge the mapping possibility in between spin-1/2 spin operators
and spin-3/2 spinful canonical Fermi operators has not been jet published in
the literature. The presented transformation allows also this fact.

I would like to finish the introductory part by a personal assessment. Often is
said, that in the majority of cases the motivation of a research is a
research gap. In the present case the motivation was not this, instead it was
a question: This transformation is it possible to be done ? Why ?
Because if it is possible, that shows that the local information contained in
the spin operators is much broader than one considers today. Since it turned
out that it is possible, this projected out new perspectives that fill up gaps
whose existence become to be clear only now. These were enumerated before at
the points i)-v).

The main results of the paper are contained in Eqs.(\ref{EQU4}, \ref{EQU5},
\ref{EQU6}, and \ref{EQU7}). The transformation can be applied for an arbitrary
Hamiltonian constructed from quantum spin-1/2 operators. For exemplification,
the case of the 1D Heisenberg Hamiltonian in Eq.(\ref{EQU8}) is analyzed here
whose transformation in locally anticommuting operators is presented in details
in Eqs.(\ref{EQU9},\ref{EQU11}-\ref{EQU20}). Since the transformation itself
contains six
independent arbitrary free parameters, a concrete application in mapping to a
two band
canonical Fermi operator Hamiltonian must be done by an appropriate fine tuning
of the free parameters. This step is also exemplified by the Hamiltonian in
Eq.(\ref{EQU36}) which describes a two band system with many-body spin-orbit
interactions possessing density dependent hoppings, a subject which is
intensively analyzed in the recent literature
\cite{JW10,JW11,JW12,JW13,JW14,JW15,JW16,JW17,JW18,JW19,JW20,JW21,JW22,JW23,
JW24,JW25,JW26,JW27,JW28}.

The remaining part of the paper is structured as follows: Section II
presents in 1D the transformation of quantum spin-1/2 operators to spinful
canonical Fermi operators present in two different bands, Section III
describes the application to the Heisenberg Hamiltonian, Section IV
explicitly presents in a fine tuned case the exact mapping to spinful canonical
Fermi operators describing two different bands, Section V presents the
summary and discussions, and Appendix A containing mathematical details,
closes the presentation.

\section{Quantum spin-1/2 operators transformed to spinful canonical Fermi
operators present in two different bands}

Let us start from quantum spin operators satisfying the  required
commutation rules, namely on-site
\begin{eqnarray} 
[\hat S^{x}_i, \hat S^{y}_i ]= i  \hat S^{z}_i, \:
[\hat S^{y}_i, \hat S^{z}_i ]= i  \hat S^{x}_i, \:
[\hat S^{z}_i, \hat S^{x}_i ]= i  \hat S^{y}_i, \:
[\hat S^{2}_i, \hat S^{\alpha}_i ]= 0,
\label{EQU1}
\end{eqnarray}
and inter-site ($i \ne j$)
\begin{eqnarray}
[\hat S^{\alpha}_i, \hat S^{\beta}_j ]= 0,
\label{EQU2}
\end{eqnarray}
where $\alpha,\beta=x,y,z$ holds.
Furthermore, in the aim to satisfy during an exact transformation the
requirements of
Eq.(\ref{EQU1}), we take 4 operators ($\hat a_i, \hat b_i, \hat e_i,
\hat d_i$), that anti-commute on-site
\begin{eqnarray}
  \{\hat g_{1,i}, \hat g^{\dagger}_{1,i}\} = \{\hat g_{2,i}, \hat g^{\dagger}_{2,i}\}
  = 1, \:
\{\hat g_{1,i}, \hat g^{\dagger}_{2,i}\} = \{\hat g^{\dagger}_{1,i}, \hat g_{2,i}\} =
\{\hat g_{1,i}, \hat g_{2,i}\} = \{\hat g^{\dagger}_{1,i}, \hat g^{\dagger}_{2,i}\}
=0, 
\label{EQU3}
\end{eqnarray}
where $\hat g_{1,i} ,\hat g_{2,i}$ can be $\hat a_i, \hat b_i, \hat e_i,
\hat d_i$, but $\hat g_{1,i} \ne \hat g_{2,i}$ holds.

After this step, taking into account that the operators  $\hat S^{\alpha}_i$ are
Hermitian, one defines
\begin{eqnarray}
&& \hat S^{x}_i = \frac{\hat a^{\dagger}_i + \hat a_i}{2X} +
\frac{\hat b^{\dagger}_i + \hat b_i}{2Y} + \frac{\hat e^{\dagger}_i + \hat e_i}{2Z}
+ \frac{\hat d^{\dagger}_i + \hat d_i}{2Q},
\nonumber\\
&& \hat S^{y}_i = \frac{\hat a^{\dagger}_i - \hat a_i}{2Ui} +
\frac{\hat b^{\dagger}_i - \hat b_i}{2Vi} + \frac{\hat e^{\dagger}_i - \hat e_i}{
2Wi}+ \frac{\hat d^{\dagger}_i - \hat d_i}{2Ri},
\nonumber\\
&&\hat S^{z}_i =-i(\hat S^x_i \hat S^y_i - \hat S^y_i \hat S^x_i),
\label{EQU4}
\end{eqnarray}
where
$X,Y,Z,Q,U,V,W,R$ are arbitrary parameters satisfying the constrains
\begin{eqnarray}
&& A = \frac{1}{X^2} +  \frac{1}{Y^2} + \frac{1}{Z^2} + \frac{1}{Q^2} = 1,
\nonumber\\ 
&& B = \frac{1}{U^2} +  \frac{1}{V^2} + \frac{1}{W^2} + \frac{1}{R^2} = 1,
\label{EQU5}
\end{eqnarray}
Based on Eqs.(\ref{EQU4},\ref{EQU5}) it can be checked, that now all required
commutation relations from Eq.(\ref{EQU1}) are satisfied, and supplementary
$\hat S^2= (A+B +AB)/4 =3/4$, hence $S=1/2$ holds.

I note here that the constrans from Eq.(\ref{EQU5}) are not introduced by
hand. They appear since otherwise the spin operators defined in
Eq.(\ref{EQU4}) do not satisfy the required commutation relations, and also
$S=1/2$ no more holds. As such $X,Y,...Q$, as redundant parameters remain
present on the fermionic side of the transformation, entering in the coupling
constants of the transformed fermionic Hamiltonian. Being 8 constants with 2
constrains, will represent 6 free parameters in the transformed Hamiltonian
that can be used in approaching the result of the transformation to the
Hamiltonian of interest in a particular study.  

Relating the results presented in Eqs.(\ref{EQU4},\ref{EQU5}), I note that
the starting point here is given by the first two lines of Eq.(\ref{EQU4})
containing the $\hat S^{x}_i, \hat S^{y}_i$ expressions. Starting from these,
using the first commutation relation from Eq.(\ref{EQU1}), we obtains the
$\hat S^z_i$ operator presented in last row of Eq.(\ref{EQU4}).
But at this moment three
further commutation relations remain in Eq.(\ref{EQU1}),
namely $[\hat S^{y}_i, \hat S^{z}_i ]= i  \hat S^{x}_i, \:
[\hat S^{z}_i, \hat S^{x}_i ]= i  \hat S^{y}_i, \:
[\hat S^{2}_i, \hat S^{\alpha}_i ]= 0,$
which also must be
satisfied. These provide the constrains from (\ref{EQU5}) and the presented
$S=1/2$ spin value.

Up to this moment we have not mentioned Eq.(\ref{EQU2}), the required
inter-site relations for spin operators,
which also must be satisfied. But for this, the on-site anticommuting operators
must become commuting operators inter-site. This is the reason why the
$\hat a_i, \hat b_i, \hat e_i, \hat d_i$ operators become hybrid operators which
anticommute on-site and commute inter-site. But using a generalized
Jordan-Wigner transformation, the hybrid $\hat a_i, \hat b_i, \hat e_i,
\hat d_i$ operators can be transformed in genuine canonical Fermi operators
$\hat c_i, \hat h_i, \hat f_i, \hat q_i$ which anticommute on-site and
inter-site as well
\begin{eqnarray}
&&\hat a_i =  exp [-i\pi \sum^{i-1}_{j=1} (\sum_{k=c,h,f,q} \hat n^{k}_j)]
\hat c_i, \quad \hat a^{\dagger}_i = \hat c^{\dagger}_i exp [i\pi \sum^{i-1}_{j=1}
(\sum_{k=c,h,f,q} \hat n^{k}_j)],
\nonumber\\
&&\hat b_i =  exp [-i\pi \sum^{i-1}_{j=1} (\sum_{k=c,h,f,q} \hat n^{k}_j)]
\hat h_i, \quad \hat b^{\dagger}_i = \hat h^{\dagger}_i exp [i\pi \sum^{i-1}_{j=1}
(\sum_{k=c,h,f,q} \hat n^{k}_j)],
\nonumber\\ 
&&\hat e_i =  exp [-i\pi \sum^{i-1}_{j=1} (\sum_{k=c,h,f,q} \hat n^{k}_j)]
\hat f_i, \quad \hat e^{\dagger}_i = \hat f^{\dagger}_i exp [i\pi \sum^{i-1}_{j=1}
(\sum_{k=c,h,f,q} \hat n^{k}_j)],
\nonumber\\
&&\hat d_i =  exp [-i\pi \sum^{i-1}_{j=1} (\sum_{k=c,h,f,q} \hat n^{k}_j)]
\hat q_i, \quad \hat d^{\dagger}_i = \hat q^{\dagger}_i exp [i\pi \sum^{i-1}_{j=1}
(\sum_{k=c,h,f,q} \hat n^{k}_j)],
\label{EQU6}
\end{eqnarray}
where $\hat n^k_j =\hat k^{\dagger}_j \hat k_j$ is the particle number operator
for the k-fermions.

Taking into consideration the anticommutation relations satisfied by the
fermionic operators $\hat c_i, \hat h_i, \hat f_i, \hat q_i$ one takes
\begin{eqnarray}
\hat c_i = \hat c_{i,\uparrow}, \quad \hat h_i = \hat c_{i,\downarrow}, \quad
\hat f_i = \hat f_{i,\uparrow}, \quad \hat q_i = \hat f_{i,\downarrow}.
\label{EQU7}  
\end{eqnarray}
In this manner, the starting quantum S=1/2 spin operators have been exactly
transformed in genuine canonical spinful Fermi operators $\hat c_{i,\sigma}$ and
$\hat f_{i,\sigma}$ describing two bands.

At this stage I note that if we would like to transform exactly the quantum
spin-1/2 spin operators in spin-1/2 canonical Fermi operators,
Eq.(\ref{EQU7})
is the unique possibility. Since the obtained relations are symmetric in
the $\hat c_i, \hat h_i, \hat f_i, \hat q_i$ operators, the choices in
Eq.(\ref{EQU7}) (e.g. $\hat h_i = \hat c_{i,\downarrow}$ and not
$\hat q_i = \hat c_{i,\downarrow}$) not diminish the generality of the presented
relations. As another potential possibility, I mention that based on the
presented transformation, the spin-1/2 quantum spin operators can be transformed
in canonical spin-3/2 Fermi operators as well, by taking $\hat c_i =
\hat c_{i,-3/2}, \hat h_i = \hat c_{i,-1/2}, \hat f_i = \hat c_{i,+1/2},
\hat q_i = \hat c_{i,+3/2}$.

Concerning the novelty of the presented transformation I mention that relative
to Refs. [1,2], i) two more hybrid operators have been used, whose total number
is presently four in Eq.(4), and ii) the Jordan-Wigner transformation in
Refs.[1,2] contained two particle number operators in the exponent, while in
Eq.(6), one has four different operators of this type. These differences are
motivated by the fact that in the final result four spinful canonical Fermi
operators were obtained, and for this four hybrid operators were needed.
The used operators on the fermionic side of the mapping are summarized in
Table 1.
\begin{table}[h!]
\centering
\begin{tabular}{|c|c|c|}
\hline
Hybrid operators & Fermi operators & Spinful Fermi Operators \\ [1ex]
\hline
\hline
$\hat a_i$ & $\hat c_i$ & $\hat c_{i,\uparrow}$ \\
\hline
$\hat b_i$ & $\hat h_i$ & $\hat c_{i,\downarrow}$ \\
\hline
$\hat e_i$ & $\hat f_i$ & $\hat f_{i,\uparrow}$ \\
\hline
$\hat d_i$ & $\hat q_i$ & $\hat f_{i,\downarrow}$ \\ [1ex]
\hline
\end{tabular}
\caption{Table presenting the used operators. A given row presents how the
operators transform from each other during the transformation.}
\label{table:1}
\end{table}    

The transformation framework (for an arbitrary Hamiltonian containing spin-1/2
operators) is as follows: In the starting spin Hamiltonian we introduce the
spin operators presented in Eq.(1), and we obtain the initial Hamiltonian in
function of the hybrid operators $a_i, b_i, e_i$, and $d_i$.
After this step, using for each term of the obtained Hamiltonian the
extended Jordan-Wigner transformation from Eq.(6) together with the
specifications from Eq.(7) we obtain the starting Hamiltonian in terms of
spinful canonical Fermi operators present in two bands. This is the final
transformed Hamiltonian, containing all redundant free parameters X,Y,..Q.

\section{The case of the Heisenberg Hamiltonian}

The presented transformation can be applied to an arbitrary S=1/2 quantum
spin Hamiltonian. Here I am exemplifying it by fully performing the
transformation from Eq.(\ref{EQU4}) in the
case of the Heisenberg Hamiltonian $\hat H = \hat H_z + \hat H_{x,y}$.
This exemplification choice is used since the Heisenberg
Hamiltonian is applied most often in the study of the spin chains.  

For the presented components in $\hat H$ one has
\begin{eqnarray}
\hat H_z = \sum_i J_z S^z_i S^z_{i+1}, \quad
\hat H_{x,y} = \sum_i (J_x S^x_i S^x_{i+1} + J_y S^y_i S^y_{i+1})
\label{EQU8}
\end{eqnarray}
In the case of transversal component one finds
$\hat H_{x,y} = \sum_i \hat H_{x,y,i}$, where
\begin{eqnarray}
\hat H_{x,y,i} &=& (\frac{J_x}{4X^2}-\frac{J_y}{4U^2})(\hat a^{\dagger}_i
\hat a^{\dagger}_{
i+1} + \hat a_i \hat a_{i+1}) + (\frac{J_x}{4X^2}+\frac{J_y}{4U^2})(\hat a^{
\dagger}_i\hat a_{i+1} + \hat a_i \hat a^{\dagger}_{i+1})
\nonumber\\
&+&(\frac{J_x}{4Y^2}-\frac{J_y}{4V^2})(\hat b^{\dagger}_i\hat b^{\dagger}_{
i+1} + \hat b_i \hat b_{i+1}) + (\frac{J_x}{4Y^2}+\frac{J_y}{4V^2})(\hat b^{
\dagger}_i\hat b_{i+1} + \hat b_i \hat b^{\dagger}_{i+1})
\nonumber\\
&+&(\frac{J_x}{4Z^2}-\frac{J_y}{4W^2})(\hat e^{\dagger}_i\hat e^{\dagger}_{
i+1} + \hat e_i \hat e_{i+1}) + (\frac{J_x}{4Z^2}+\frac{J_y}{4W^2})(\hat e^{
\dagger}_i\hat e_{i+1} + \hat e_i \hat e^{\dagger}_{i+1})
\nonumber\\
&+&(\frac{J_x}{4Q^2}-\frac{J_y}{4R^2})(\hat d^{\dagger}_i\hat d^{\dagger}_{
i+1} + \hat d_i \hat d_{i+1}) + (\frac{J_x}{4Q^2}+\frac{J_y}{4R^2})(\hat d^{
\dagger}_i\hat d_{i+1} + \hat d_i \hat d^{\dagger}_{i+1})
\nonumber\\
&+&(\frac{J_x}{4XY}-\frac{J_y}{4UV})(\hat a^{\dagger}_i\hat b^{\dagger}_{i+1} +
\hat a_i \hat b_{i+1} + \hat b^{\dagger}_i \hat a^{\dagger}_{i+1} + \hat b_i
\hat a_{i+1})
\nonumber\\
&+& (\frac{J_x}{4XY}+\frac{J_y}{4UV})(\hat a^{\dagger}_i\hat b_{i+1} + \hat a_i
\hat b^{\dagger}_{i+1}  + \hat b^{\dagger}_i \hat a_{i+1} + \hat b_i
\hat a^{\dagger}_{i+1})
\nonumber\\
&+&(\frac{J_x}{4XZ}-\frac{J_y}{4UW})(\hat a^{\dagger}_i\hat e^{\dagger}_{i+1} +
\hat a_i \hat e_{i+1} + \hat e^{\dagger}_i \hat a^{\dagger}_{i+1} + \hat e_i
\hat a_{i+1})
\nonumber\\
&+& (\frac{J_x}{4XZ}+\frac{J_y}{4UW})(\hat a^{\dagger}_i\hat e_{i+1} + \hat a_i
\hat e^{\dagger}_{i+1}  + \hat e^{\dagger}_i \hat a_{i+1} + \hat e_i
\hat a^{\dagger}_{i+1})
\nonumber\\
&+&(\frac{J_x}{4XQ}-\frac{J_y}{4UR})(\hat a^{\dagger}_i\hat d^{\dagger}_{i+1} +
\hat a_i \hat d_{i+1} + \hat d^{\dagger}_i \hat a^{\dagger}_{i+1} + \hat d_i
\hat a_{i+1})
\nonumber\\
&+& (\frac{J_x}{4XQ}+\frac{J_y}{4UR})(\hat a^{\dagger}_i\hat d_{i+1} + \hat a_i
\hat d^{\dagger}_{i+1}  + \hat d^{\dagger}_i \hat a_{i+1} + \hat d_i
\hat a^{\dagger}_{i+1})
\nonumber\\
&+&(\frac{J_x}{4YZ}-\frac{J_y}{4VW})(\hat b^{\dagger}_i\hat e^{\dagger}_{i+1} +
\hat b_i \hat e_{i+1} + \hat e^{\dagger}_i \hat b^{\dagger}_{i+1} + \hat e_i
\hat b_{i+1})
\nonumber\\
&+& (\frac{J_x}{4YZ}+\frac{J_y}{4VW})(\hat b^{\dagger}_i\hat e_{i+1} + \hat b_i
\hat e^{\dagger}_{i+1}  + \hat e^{\dagger}_i \hat b_{i+1} + \hat e_i
\hat b^{\dagger}_{i+1})
\nonumber\\
&+&(\frac{J_x}{4YQ}-\frac{J_y}{4VR})(\hat b^{\dagger}_i\hat d^{\dagger}_{i+1} +
\hat b_i \hat d_{i+1} + \hat d^{\dagger}_i \hat b^{\dagger}_{i+1} + \hat d_i
\hat b_{i+1})
\nonumber\\
&+& (\frac{J_x}{4YQ}+\frac{J_y}{4VR})(\hat b^{\dagger}_i\hat d_{i+1} + \hat b_i
\hat d^{\dagger}_{i+1}  + \hat d^{\dagger}_i \hat b_{i+1} + \hat d_i
\hat b^{\dagger}_{i+1})
\nonumber\\
&+&(\frac{J_x}{4ZQ}-\frac{J_y}{4WR})(\hat e^{\dagger}_i\hat d^{\dagger}_{i+1} +
\hat e_i \hat d_{i+1} + \hat d^{\dagger}_i \hat e^{\dagger}_{i+1} + \hat d_i
\hat e_{i+1})
\nonumber\\
&+& (\frac{J_x}{4ZQ}+\frac{J_y}{4WR})(\hat e^{\dagger}_i\hat d_{i+1} + \hat e_i
\hat d^{\dagger}_{i+1}  + \hat d^{\dagger}_i \hat e_{i+1} + \hat d_i
\hat e^{\dagger}_{i+1}).
\label{EQU9}
\end{eqnarray}

In the case of $\hat H_z$ one obtains
\begin{eqnarray}
\hat H_z = J_z \sum_i \sum_{n=1}^{10} \hat P_{n,i}.
\label{EQU10}
\end{eqnarray}
All contributions from Eq.(\ref{EQU10}) are presented in Appendix A.

The results presented in Eqs.(\ref{EQU9},\ref{EQU11}-\ref{EQU20}) have been
simply obtained by introducing Eq.(\ref{EQU4}) in Eq.(\ref{EQU8}),
effectuating the products $\hat S^{\alpha}_i \hat S^{\alpha}_{i+1}$ and adding
all obtained contributions.

\section{Possible applications}

In order to exemplify the transformation to spinful canonical Fermi operators
which are belong to two bands, a concrete application is presented by
effectuating the transformation from Eqs.(\ref{EQU6},\ref{EQU7})  to
the Hamiltonian from Eqs.(\ref{EQU9},\ref{EQU11}-\ref{EQU20}) in a fine tuned
case. This
was chosen to be
$J_z=0, J_x=J_y$ and
\begin{eqnarray}
\frac{V}{Y}=\frac{W}{Z}=\frac{R}{Q}=\frac{U}{X} = \alpha,
\label{EQU35}
\end{eqnarray}
where $\alpha^2=1$ (necessary in order to satisfy (\ref{EQU5})). In order to
enhance the presentation of the results I also introduce the density dependent
operator
\begin{eqnarray}
\hat N_{k_1,\sigma;k_2}=(1-2 \hat n^{k_1}_{i,\sigma})
(1-2 \hat n^{k_2}_{i,\uparrow})(1-2 \hat n^{k_2}_{i,\downarrow}),
\label{EQU21}
\end{eqnarray}
where $k_1,k_2 = c,f$, the $\hat N_{k_1,\sigma;k_2}$ term being defined
always at the site $i$, and $\sigma=\uparrow,\downarrow$ holds, respectively. 
The density dependent operator from Eq.(\ref{EQU21}) appears during the
transformation of the hybrid operators in spinful canonical Fermi operators
via the Jordan-Wigner transformation from Eq.(\ref{EQU6},\ref{EQU7}). For
example
\begin{eqnarray}
\hat a^{\dagger}_i \hat a_{i+1} = \hat c^{\dagger}_{i,\uparrow}
\hat c_{i+1,\uparrow}  \hat N_{c,\downarrow;f}, \quad
\hat b^{\dagger}_i \hat b_{i+1} = \hat c^{\dagger}_{i,\downarrow}
\hat c_{i+1,\downarrow}  \hat N_{c,\uparrow;f}, \: \: etc.
\label{EQU21a}
\end{eqnarray}

The result of the transformation becomes
\begin{eqnarray}
\hat H &=& \frac{J_x}{2} \sum_i \{ (\hat c^{\dagger}_{i,\uparrow}
\hat c_{i+1,\uparrow}
+ H.c.)\frac{\hat N(c,\downarrow;f)}{X^2} + (\hat c^{\dagger}_{i,\downarrow}
\hat c_{i+1,\downarrow} + H.c.)\frac{\hat N(c,\uparrow;f)}{Y^2}
\nonumber\\
&+&
(\hat f^{\dagger}_{i,\uparrow} \hat f_{i+1,\uparrow} + H.c.)
\frac{\hat N(f,\downarrow;c)}{Z^2} +
(\hat f^{\dagger}_{i,\downarrow}\hat f_{i+1,\downarrow} + H.c.)
\frac{\hat N(f,\uparrow;c)}{Q^2}
\nonumber\\
&+&(\hat c^{\dagger}_{i,\uparrow}
\hat f_{i+1,\uparrow} + H.c.)\frac{\hat N(c,\downarrow;f)}{XZ} +
(\hat f^{\dagger}_{i,\uparrow}
\hat c_{i+1,\uparrow} + H.c.)\frac{\hat N(f,\downarrow;f)}{XZ}
\nonumber\\
&+&(\hat c^{\dagger}_{i,\downarrow}
\hat f_{i+1,\downarrow} + H.c.)\frac{\hat N(c,\uparrow;f)}{YQ} +
(\hat f^{\dagger}_{i,\downarrow}
\hat c_{i+1,\downarrow} + H.c.)\frac{\hat N(f,\uparrow;c)}{YQ}
\nonumber\\
&+& \sum_{\sigma} [(\hat c^{\dagger}_{i,\sigma}
\hat c_{i+1,-\sigma} + H.c.)\frac{\hat N(c,-\sigma;f)}{XY} +
(\hat f^{\dagger}_{i,\sigma}\hat f_{i+1,-\sigma} + H.c.)
\frac{\hat N(f,-\sigma;c)}{ZQ}]
\nonumber\\
&+&(\hat c^{\dagger}_{i,\uparrow}
\hat f_{i+1,\downarrow} + H.c.)\frac{\hat N(c,\downarrow;f)}{XQ} +
(\hat f^{\dagger}_{i,\downarrow}
\hat c_{i+1,\uparrow} + H.c.)\frac{\hat N(f,\uparrow;c)}{XQ}
\nonumber\\
&+&(\hat c^{\dagger}_{i,\downarrow}
\hat f_{i+1,\uparrow} + H.c.)\frac{\hat N(c,\uparrow;f)}{YZ} +
(\hat f^{\dagger}_{i,\uparrow}
\hat c_{i+1,\downarrow} + H.c.)\frac{\hat N(f,\downarrow;c)}{YZ} \}.
\label{EQU36}  
\end{eqnarray}
Note that the Hamiltonian from (\ref{EQU36}) is perfectly
equivalent to the 1D quantum spin-1/2 Heisenberg Hamiltonian
$\hat H = J_x \sum_i(\hat S^x_i \hat S^x_{i+1} + \hat S^y_i \hat S^y_{i+1})$,
see Eq.(\ref{EQU8}).

I note that applying the Jordan-Wigner transformation in obtaining the final
transformed Hamiltonian from (\ref{EQU36}), the Hamiltonian Eq.(\ref{EQU8})
was considered with open boundary conditions. I note that it is also possible
to extend the transformation to Hamiltonians taken with periodic boundary
conditions as well on the line of Ref.\cite{JW4}.
 
Relating the use of Eqs.(\ref{EQU6},\ref{EQU7}) in transforming the hybrid
$\hat a_i,\hat b_i,\hat e_i,\hat d_i$ operators present in
Eqs.(\ref{EQU9}, \ref{EQU11}-\ref{EQU20}) in canonical Fermi
operators presented in Eq.(\ref{EQU7}),
extreme details have been provided in Ref.[\cite{JW2}], where the use of a
generalized Jordan-Wigner transformation was explained in details.
Summarizing the
presented technique, three main properties
should be taken into account, namely: a) Given by the canonical Fermi nature of
the operators $\hat c_{i,\sigma},\hat f_{i,\sigma}$, all operators
$\hat n^{\mu}_{i,\sigma}$ (where $\mu=c,f$) commute for all $\mu$, all $\sigma$,
and all sites $i$; b) One has $exp[\pm i\pi \hat n ] =1-2 \hat n$, and 
$\exp[\pm 2i\pi \hat n]=1$ since in the fermionic case, the
particle number operator $\hat n$ has only eigenvalues zero or one. c) Taking
into account the first relation from b), multiplying $exp[\pm i\pi \hat n ]$
from left or right by a fermionic operator, a sign difference appears, e.g.
for $\hat n=\hat c^{\dagger} \hat c$, one has
$\hat c^{\dagger}exp[\pm i\pi \hat n ]=\hat c^{\dagger}$, while
$exp[\pm i\pi \hat n ]\hat c^{\dagger}=-\hat c^{\dagger}$. Similarly
$\hat c \: exp[\pm i\pi \hat n ]= - \hat c$, while
$exp[\pm i\pi \hat n ]\hat c=\hat c$, etc. In this manner one obtains
the Hamiltonian from Eq.(\ref{EQU36}).

When we apply the described transformation, we transform a given spin
Hamiltonian and we obtain the transformed Hamiltonian written in
fermionic language. This transformed Hamiltonian contains 6 free parameters
which are arbitrary. How we chose these parameters, depends on what kind of
Hamiltonian is in our focus on the fermionic side of the mapping. For example,
in the presented case of Eq.(14), the choice Eq.(11) has been used in order
to obtain a Hamiltonian describing a system with SOC, also under the action of
correlated hopping which is intensively studied in the literature. As seen in
Eq.(14) the fine tuning cuts some free parameters, and introduce the remaining
parameters in the coupling constants of the transformed Hamiltonian. 

The Hamiltonian from (\ref{EQU36}) describes in general a two band system in
1D which has a kinetic energy (hopping terms without spin-flip), interacts
with an external magnetic field via a Zeeman term (spin-up and spin-down hopping
terms may not have the same amplitude), and contains many-body spin-orbit 
coupling (SOC) (providing the spin-flip hopping terms). Besides, all hopping
terms there are density dependent (this characteristic being also called
bond-charge interaction or ``correlated hopping'', see e.g. Ref.[\cite{JW9}]).

The presence of SOC in 1D systems is intensively analyzed
\cite{JW10,JW11,JW12}, and the majority of systems of interest are multiband
systems \cite{JW13,JW14,JW15,JW16}. In many-body itinerant systems SOC produces
usually spin-flip hoppings (see e.g. Ref.(\cite{JW15})). On this line, relating
the hopping terms especially in strongly correlated systems, it turns out that
these contributions taken as correlated hoppings, the system characteristics
can be reproduced much better. Hence this possibility is intensively analyzed
nowadays \cite{JW17,JW18,JW19,JW20}. Consequently, it is not surprising, that
recent SOC studies are often made entangled with density dependent hopping
contributions \cite{JW21,JW22,JW23,JW24,JW25,JW26}, even the interdependence
of SOC and correlations has been pointed out \cite{JW27,JW28}.
This is the field in which the Hamiltonian from Eq.(\ref{EQU36}) can provide
valuable, exact information, that can guide the research and description
possibilities in the direction of the density dependent hoppings in the presence
of SOC couplings. Nevertheless, since the fermionic side of the mapping is
extremely rich, a substantial amount of work is necessary in order to use
properly the fine tuning in the direction of a desired fermionic system.

\section{Summary and Discussions}

Starting from the aim to possess exact mappings in between spin models and
itinerant fermion models, quantum spin-1/2 spin operators were transformed not
only in spinless Fermi operators \cite{JW4}, but also in spinful canonical
Fermi operators both in 1D \cite{JW1} and 2D \cite{JW2}, opening the doors for
the study of exact interconnections in between spin-1/2 quantum spin models and
realistic (i.e. spinful) itinerant spin-1/2 fermion models. On this background,
in the presented paper it is demonstrated in 1D, that spin-1/2 quantum spin
operators can be also exactly transformed in canonical spinful spin-1/2 Fermi
operators present in two bands. For exemplification, the transformation is
presented in the case of the Heisenberg Hamiltonian. Since the operator
transformation itself contains six free and independent parameters, the mapping
in between the spin system, and the itinerant two bands spinful Fermi system,
in a concrete case, must be done by a fine tuning of the arbitrary parameters.
This step is also exemplified in the case of a two band system, containing
many-body spin-orbit interactions possessing density dependent hoppings,
of large interest today \cite{JW21,JW22,JW23,JW24,JW25,JW26,JW27,JW28}.

I further note, that the presented transformation can also be used in the exact
transformation of spin-1/2 quantum spin operators in spinful spin-3/2 canonical
Fermi operators. As a conjecture, I mention that spin-1/2 quantum spin
operators can be exactly transformed in spinful spin-1/2 canonical Fermi
operators describing fermions present in an arbitrary number of bands. For
this, only the number of operators in the first two lines of Eq.(\ref{EQU4})
should be increased, maintaining a priory the $\hat S^x_i,\hat S^y_i$ spin
component operators Hermitian. By this procedure is also possible to map the
spin-1/2 spin operators in spinful canonical Fermi operators holding
spin-(2n+1)/2.

For readers that would like to perform numerical calculations in this
direction, I am mentioning that in order to perform the transformation
i) First is necessary to have a package that can manipulate anticommuting
operators; ii) After this step, the expressions of the spin operators from
Eq.(4) can be introduced in the analyzed spin Hamiltonian $\hat H$, obtaining
the expression of $\hat H$ in terms of the hybrid operators; iii) Now is
necessary to have a package that is able to provide the Jordan-Wigner
transformation of an arbitrary contribution written in terms of hybrid
operators. This must be constructed based on Eqs.(6,7) and the rules
presented on bottom pg.10 to top pg.11. iv) Finally, based on the package
constructed at iii), all individual terms from $\hat H$, term by term, must
be transformed in spinful canonical Fermi operators.



ACKNOWLEDGEMENT:

Research supported by Chinese Academy of Sciences President's International
Fellowship Initiative, PIFI Grant No: 2025PVA0087.

E-MAIL ADDRESS OF THE AUTHOR:

gulacsi@phys.unideb.hu

DATA AVAILABILITY STATEMENT:

The data that support the results are available from the author upon
reasonable request.



\appendix

\section{The individual contributions in $\hat H_z$}

This Appendix collects the contributions from Eq.(\ref{EQU10})

\begin{eqnarray}
\hat P_{1,i} &=& \frac{1}{4}(\frac{1}{XU}+\frac{1}{YV}+\frac{1}{ZW}+\frac{1}{QR}
)^2 -\frac{1}{2}(\frac{1}{XU}+\frac{1}{YV}+\frac{1}{ZW}+\frac{1}{QR})
(\frac{\hat n^a_i+\hat n^a_{i+1}}{XU} + \frac{\hat n^b_i+\hat n^b_{i+1}}{YV}
\nonumber\\
&+& \frac{\hat n^e_i+\hat n^e_{i+1}}{ZW} + \frac{\hat n^d_i+\hat n^d_{i+1}}{QR})
+(\frac{\hat n^a_i}{XU}+\frac{\hat n^b_i}{YV}+\frac{\hat n^e_i}{ZW}+
\frac{\hat n^d_i}{QR})(\frac{\hat n^a_{i+1}}{XU}+\frac{\hat n^b_{i+1}}{YV}+
\frac{\hat n^e_{i+1}}{ZW}+\frac{\hat n^d_{i+1}}{QR})
\nonumber\\
&-&\frac{1}{4}(\frac{1}{XU}+\frac{1}{YV}+\frac{1}{ZW}+\frac{1}{QR})[
(\frac{1}{UY}-\frac{1}{XV})(\hat a^{\dagger}_i \hat b^{\dagger}_i + \hat b_i
\hat a_i + \hat a^{\dagger}_{i+1} \hat b^{\dagger}_{i+1} + \hat b_{i+1} \hat a_{i+1})
\nonumber\\
+(\frac{1}{ZU}&-&\frac{1}{XW})(\hat a^{\dagger}_i \hat e^{\dagger}_i + \hat e_i
\hat a_i + \hat a^{\dagger}_{i+1} \hat e^{\dagger}_{i+1} + \hat e_{i+1} \hat a_{i+1})
+(\frac{1}{QU}-\frac{1}{XR})(\hat a^{\dagger}_i \hat d^{\dagger}_i + \hat d_i
\hat a_i + \hat a^{\dagger}_{i+1} \hat d^{\dagger}_{i+1} + \hat d_{i+1} \hat a_{i+1})
\nonumber\\
+(\frac{1}{ZV}&-&\frac{1}{YW})(\hat b^{\dagger}_i \hat e^{\dagger}_i + \hat e_i
\hat b_i + \hat b^{\dagger}_{i+1} \hat e^{\dagger}_{i+1} + \hat e_{i+1} \hat b_{i+1})
+(\frac{1}{QV}-\frac{1}{YR})(\hat b^{\dagger}_i \hat d^{\dagger}_i + \hat d_i
\hat b_i + \hat b^{\dagger}_{i+1} \hat d^{\dagger}_{i+1} + \hat d_{i+1} \hat b_{i+1})
\nonumber\\
+(\frac{1}{QW}&-&\frac{1}{ZR})(\hat e^{\dagger}_i \hat d^{\dagger}_i + \hat d_i
\hat e_i + \hat e^{\dagger}_{i+1} \hat d^{\dagger}_{i+1} + \hat d_{i+1} \hat e_{i+1})
-(\frac{1}{XV}+\frac{1}{UY})(\hat a_i \hat b^{\dagger}_i + \hat b_i
\hat a^{\dagger}_i + \hat a_{i+1} \hat b^{\dagger}_{i+1} + \hat b_{i+1}
\hat a^{\dagger}_{i+1})
\nonumber\\%
-(\frac{1}{XW}&+&\frac{1}{ZU})(\hat a_i \hat e^{\dagger}_i + \hat e_i
\hat a^{\dagger}_i + \hat a_{i+1} \hat e^{\dagger}_{i+1} + \hat e_{i+1}
\hat a^{\dagger}_{i+1})-(\frac{1}{XR}+\frac{1}{UQ})(\hat a_i \hat d^{\dagger}_i +
\hat d_i \hat a^{\dagger}_i + \hat a_{i+1} \hat d^{\dagger}_{i+1} + \hat d_{i+1}
\hat a^{\dagger}_{i+1})
\nonumber\\
-(\frac{1}{YW}&+&\frac{1}{ZV})(\hat b_i \hat e^{\dagger}_i + \hat e_i
\hat b^{\dagger}_i + \hat b_{i+1} \hat e^{\dagger}_{i+1} + \hat e_{i+1}
\hat b^{\dagger}_{i+1})-(\frac{1}{YR}+\frac{1}{VQ})(\hat b_i \hat d^{\dagger}_i +
\hat d_i \hat b^{\dagger}_i + \hat b_{i+1} \hat d^{\dagger}_{i+1} + \hat d_{i+1}
\hat b^{\dagger}_{i+1})
\nonumber\\
-(\frac{1}{ZR}&+&\frac{1}{QW})(\hat e_i \hat d^{\dagger}_i + \hat d_i
\hat e^{\dagger}_i + \hat e_{i+1} \hat d^{\dagger}_{i+1} + \hat d_{i+1}
\hat e^{\dagger}_{i+1})].
\label{EQU11}
\end{eqnarray}
\begin{eqnarray}
&&\hat P_{2,i}= \frac{1}{2} [ (\frac{1}{UY}-\frac{1}{XV})(\hat a^{\dagger}_i
\hat b^{\dagger}_i + \hat b_i \hat a_i) + (\frac{1}{ZU}-\frac{1}{XW})
(\hat a^{\dagger}_i \hat e^{\dagger}_i + \hat e_i \hat a_i) +
(\frac{1}{QU}-\frac{1}{XR})(\hat a^{\dagger}_i
\hat d^{\dagger}_i + \hat d_i \hat a_i)
\nonumber\\
&&+ (\frac{1}{ZV}-\frac{1}{YW})(\hat b^{\dagger}_i
\hat e^{\dagger}_i + \hat e_i \hat b_i) +
(\frac{1}{QV}-\frac{1}{YR})(\hat b^{\dagger}_i
\hat d^{\dagger}_i + \hat d_i \hat b_i) +
(\frac{1}{QW}-\frac{1}{ZR})(\hat e^{\dagger}_i
\hat d^{\dagger}_i + \hat d_i \hat e_i)
\nonumber\\
&&-(\frac{1}{XV}+\frac{1}{UY})(\hat a_i \hat b^{\dagger}_i +
\hat b_i \hat a^{\dagger}_i) -
(\frac{1}{XW}+\frac{1}{UZ})(\hat a_i \hat e^{\dagger}_i +
\hat e_i \hat a^{\dagger}_i) -
(\frac{1}{XR}+\frac{1}{UQ})(\hat a_i \hat d^{\dagger}_i +
\hat d_i \hat a^{\dagger}_i)
\nonumber\\
&&-(\frac{1}{YW}+\frac{1}{ZV})(\hat b_i \hat e^{\dagger}_i +
\hat e_i \hat b^{\dagger}_i) -
(\frac{1}{YR}+\frac{1}{QV})(\hat b_i \hat d^{\dagger}_i +
\hat d_i \hat b^{\dagger}_i) -
(\frac{1}{ZR}+\frac{1}{QW})(\hat e_i \hat d^{\dagger}_i +
\hat d_i \hat e^{\dagger}_i) ]
\nonumber\\%
&&\times  (\frac{\hat n^a_{i+1}}{XU}+\frac{\hat n^b_{i+1}}{YV}+
\frac{\hat n^e_{i+1}}{ZW}+\frac{\hat n^d_{i+1}}{QR}) + \frac{1}{2}(
\frac{\hat n^a_i}{XU}+\frac{\hat n^b_i}{YV}+\frac{\hat n^e_i}{ZW}+
\frac{\hat n^d_i}{QR}) [ (\frac{1}{UY}-\frac{1}{XV})(\hat a^{\dagger}_{i+1}
\hat b^{\dagger}_{i+1} + \hat b_{i+1} \hat a_{i+1})
\nonumber\\
&&+ (\frac{1}{ZU}-\frac{1}{XW})
(\hat a^{\dagger}_{i+1} \hat e^{\dagger}_{i+1} + \hat e_{i+1} \hat a_{i+1}) +
(\frac{1}{QU}-\frac{1}{XR})(\hat a^{\dagger}_{i+1}
\hat d^{\dagger}_{i+1} + \hat d_{i+1} \hat a_{i+1})
\nonumber\\
&&+ (\frac{1}{ZV}-\frac{1}{YW})(\hat b^{\dagger}_{i+1}
\hat e^{\dagger}_{i+1} + \hat e_{i+1} \hat b_{i+1}) +
(\frac{1}{QV}-\frac{1}{YR})(\hat b^{\dagger}_{i+1}
\hat d^{\dagger}_{i+1} + \hat d_{i+1} \hat b_{i+1})
\nonumber\\
&&+ (\frac{1}{QW}-\frac{1}{ZR})(\hat e^{\dagger}_{i+1}
\hat d^{\dagger}_{i+1} + \hat d_{i+1} \hat e_{i+1})
-(\frac{1}{XV}+\frac{1}{UY})(\hat a_{i+1} \hat b^{\dagger}_{i+1} +
\hat b_{i+1} \hat a^{\dagger}_{i+1})
\nonumber\\
&&-(\frac{1}{XW}+\frac{1}{UZ})(\hat a_{i+1} \hat e^{\dagger}_{i+1} +
\hat e_{i+1} \hat a^{\dagger}_{i+1}) -
(\frac{1}{XR}+\frac{1}{UQ})(\hat a_{i+1} \hat d^{\dagger}_{i+1} +
\hat d_{i+1} \hat a^{\dagger}_{i+1})
\nonumber\\
&&-(\frac{1}{YW}+\frac{1}{ZV})(\hat b_{i+1} \hat e^{\dagger}_{i+1} +
\hat e_{i+1} \hat b^{\dagger}_{i+1}) -
(\frac{1}{YR}+\frac{1}{QV})(\hat b_{i+1} \hat d^{\dagger}_{i+1} +
\hat d_{i+1} \hat b^{\dagger}_{i+1})
\nonumber\\
&&-(\frac{1}{ZR}+\frac{1}{QW})(\hat e_{i+1} \hat d^{\dagger}_{i+1} +
\hat d_{i+1} \hat e^{\dagger}_{i+1}) ].
\label{EQU12}
\end{eqnarray}
\begin{eqnarray}
\hat P_{3,i} &=& \frac{1}{4} [ (\frac{1}{UY}-\frac{1}{XV})^2 (\hat a^{\dagger}_i
\hat b^{\dagger}_i \hat a^{\dagger}_{i+1} \hat b^{\dagger}_{i+1} +
\hat a^{\dagger}_i \hat b^{\dagger}_i \hat b_{i+1} \hat a_{i+1} +
\hat b_i \hat a_i \hat a^{\dagger}_{i+1} \hat b^{\dagger}_{i+1} +
\hat b_i \hat a_i \hat b_{i+1} \hat a_{i+1})
\nonumber\\
&+&(\frac{1}{ZU}-\frac{1}{XW})^2 (\hat a^{\dagger}_i
\hat e^{\dagger}_i \hat a^{\dagger}_{i+1} \hat e^{\dagger}_{i+1} +
\hat a^{\dagger}_i \hat e^{\dagger}_i \hat e_{i+1} \hat a_{i+1} +
\hat e_i \hat a_i \hat a^{\dagger}_{i+1} \hat e^{\dagger}_{i+1} +
\hat e_i \hat a_i \hat e_{i+1} \hat a_{i+1})
\nonumber\\
&+&(\frac{1}{QU}-\frac{1}{XR})^2 (\hat a^{\dagger}_i
\hat d^{\dagger}_i \hat a^{\dagger}_{i+1} \hat d^{\dagger}_{i+1} +
\hat a^{\dagger}_i \hat d^{\dagger}_i \hat d_{i+1} \hat a_{i+1} +
\hat d_i \hat a_i \hat a^{\dagger}_{i+1} \hat d^{\dagger}_{i+1} +
\hat d_i \hat a_i \hat d_{i+1} \hat a_{i+1})
\nonumber\\
&+&(\frac{1}{ZV}-\frac{1}{YW})^2 (\hat b^{\dagger}_i
\hat e^{\dagger}_i \hat b^{\dagger}_{i+1} \hat e^{\dagger}_{i+1} +
\hat b^{\dagger}_i \hat e^{\dagger}_i \hat e_{i+1} \hat b_{i+1} +
\hat e_i \hat b_i \hat b^{\dagger}_{i+1} \hat e^{\dagger}_{i+1} +
\hat e_i \hat b_i \hat e_{i+1} \hat b_{i+1})
\nonumber\\%
&+&(\frac{1}{QV}-\frac{1}{YR})^2 (\hat b^{\dagger}_i
\hat d^{\dagger}_i \hat b^{\dagger}_{i+1} \hat d^{\dagger}_{i+1} +
\hat b^{\dagger}_i \hat d^{\dagger}_i \hat d_{i+1} \hat b_{i+1} +
\hat d_i \hat b_i \hat b^{\dagger}_{i+1} \hat d^{\dagger}_{i+1} +
\hat d_i \hat b_i \hat d_{i+1} \hat b_{i+1})
\nonumber\\
&+&(\frac{1}{QW}-\frac{1}{ZR})^2 (\hat e^{\dagger}_i
\hat d^{\dagger}_i \hat e^{\dagger}_{i+1} \hat d^{\dagger}_{i+1} +
\hat e^{\dagger}_i \hat d^{\dagger}_i \hat d_{i+1} \hat e_{i+1} +
\hat d_i \hat e_i \hat e^{\dagger}_{i+1} \hat d^{\dagger}_{i+1} +
\hat d_i \hat e_i \hat d_{i+1} \hat e_{i+1})
\nonumber\\
&+&(\frac{1}{XV}+\frac{1}{UY})^2 (\hat a_i
\hat b^{\dagger}_i \hat a_{i+1} \hat b^{\dagger}_{i+1} +
\hat a_i \hat b^{\dagger}_i \hat b_{i+1} \hat a^{\dagger}_{i+1} +
\hat b_i \hat a^{\dagger}_i \hat a_{i+1} \hat b^{\dagger}_{i+1} +
\hat b_i \hat a^{\dagger}_i \hat b_{i+1} \hat a^{\dagger}_{i+1})
\nonumber\\
&+&(\frac{1}{XW}+\frac{1}{ZU})^2 (\hat a_i
\hat e^{\dagger}_i \hat a_{i+1} \hat e^{\dagger}_{i+1} +
\hat a_i \hat e^{\dagger}_i \hat e_{i+1} \hat a^{\dagger}_{i+1} +
\hat e_i \hat a^{\dagger}_i \hat a_{i+1} \hat e^{\dagger}_{i+1} +
\hat e_i \hat a^{\dagger}_i \hat e_{i+1} \hat a^{\dagger}_{i+1})
\nonumber\\
&+&(\frac{1}{XR}+\frac{1}{QU})^2 (\hat a_i
\hat d^{\dagger}_i \hat a_{i+1} \hat d^{\dagger}_{i+1} +
\hat a_i \hat d^{\dagger}_i \hat d_{i+1} \hat a^{\dagger}_{i+1} +
\hat d_i \hat a^{\dagger}_i \hat a_{i+1} \hat d^{\dagger}_{i+1} +
\hat d_i \hat a^{\dagger}_i \hat d_{i+1} \hat a^{\dagger}_{i+1})
\nonumber\\
&+&(\frac{1}{YW}+\frac{1}{ZV})^2 (\hat b_i
\hat e^{\dagger}_i \hat b_{i+1} \hat e^{\dagger}_{i+1} +
\hat b_i \hat e^{\dagger}_i \hat e_{i+1} \hat b^{\dagger}_{i+1} +
\hat e_i \hat b^{\dagger}_i \hat b_{i+1} \hat e^{\dagger}_{i+1} +
\hat e_i \hat b^{\dagger}_i \hat e_{i+1} \hat b^{\dagger}_{i+1})
\nonumber\\
&+&(\frac{1}{YR}+\frac{1}{QV})^2 (\hat b_i
\hat d^{\dagger}_i \hat b_{i+1} \hat d^{\dagger}_{i+1} +
\hat b_i \hat d^{\dagger}_i \hat d_{i+1} \hat b^{\dagger}_{i+1} +
\hat d_i \hat b^{\dagger}_i \hat b_{i+1} \hat d^{\dagger}_{i+1} +
\hat d_i \hat b^{\dagger}_i \hat d_{i+1} \hat b^{\dagger}_{i+1})
\nonumber\\
&+&(\frac{1}{ZR}+\frac{1}{QW})^2 (\hat e_i
\hat d^{\dagger}_i \hat e_{i+1} \hat d^{\dagger}_{i+1} +
\hat e_i \hat d^{\dagger}_i \hat d_{i+1} \hat e^{\dagger}_{i+1} +
\hat d_i \hat e^{\dagger}_i \hat e_{i+1} \hat d^{\dagger}_{i+1} +
\hat d_i \hat e^{\dagger}_i \hat d_{i+1} \hat e^{\dagger}_{i+1})].
\label{EQU13}
\end{eqnarray}
\begin{eqnarray}
\hat P_{4,i} &=& \frac{1}{4} \{ (\frac{1}{UY}-\frac{1}{XV}) [ (\frac{1}{ZU}-
\frac{1}{XW})(\hat a^{\dagger}_i \hat b^{\dagger}_i \hat a^{\dagger}_{i+1}
\hat e^{\dagger}_{i+1} + \hat a^{\dagger}_i \hat b^{\dagger}_i \hat e_{i+1}
\hat a_{i+1} + \hat b_i \hat a_i \hat a^{\dagger}_{i+1} \hat e^{\dagger}_{i+1}
+ \hat b_i \hat a_i \hat e_{i+1} \hat a_{i+1}
\nonumber\\
&+& \hat a^{\dagger}_i \hat e^{\dagger}_i
\hat a^{\dagger}_{i+1} \hat b^{\dagger}_{i+1} + \hat a^{\dagger}_i \hat e^{\dagger}_i
\hat b_{i+1} \hat a_{i+1} + \hat e_i \hat a_i \hat a^{\dagger}_{i+1}
\hat b^{\dagger}_{i+1} + \hat e_i \hat a_i \hat b_{i+1} \hat a_{i+1}) +
(\frac{1}{UQ}-\frac{1}{XR})(\hat a^{\dagger}_i \hat b^{\dagger}_i
\hat a^{\dagger}_{i+1} \hat d^{\dagger}_{i+1}
\nonumber\\
&+& \hat a^{\dagger}_i \hat b^{\dagger}_i
\hat d_{i+1} \hat a_{i+1} + \hat b_i \hat a_i \hat a^{\dagger}_{i+1}
\hat d^{\dagger}_{i+1} + \hat b_i \hat a_i \hat d_{i+1} \hat a_{i+1}
+ \hat a^{\dagger}_i \hat d^{\dagger}_i
\hat a^{\dagger}_{i+1} \hat b^{\dagger}_{i+1} + \hat a^{\dagger}_i \hat d^{\dagger}_i
\hat b_{i+1} \hat a_{i+1} + \hat d_i \hat a_i \hat a^{\dagger}_{i+1}
\hat b^{\dagger}_{i+1}
\nonumber\\
&+& \hat d_i \hat a_i \hat b_{i+1} \hat a_{i+1}) + (\frac{1}{ZV}-\frac{1}{YW})
(\hat a^{\dagger}_i \hat b^{\dagger}_i\hat b^{\dagger}_{i+1} \hat e^{\dagger}_{i+1}+
\hat a^{\dagger}_i \hat b^{\dagger}_i \hat e_{i+1} \hat b_{i+1} +
\hat b_i \hat a_i \hat b^{\dagger}_{i+1} \hat e^{\dagger}_{i+1} +
\hat b_i \hat a_i \hat e_{i+1} \hat b_{i+1}
\nonumber\\%
&+&\hat b^{\dagger}_i \hat e^{\dagger}_i\hat a^{\dagger}_{i+1} \hat b^{\dagger}_{i+1}+
\hat b^{\dagger}_i \hat e^{\dagger}_i \hat b_{i+1} \hat a_{i+1} +
\hat e_i \hat b_i \hat a^{\dagger}_{i+1} \hat b^{\dagger}_{i+1} +
\hat e_i \hat b_i \hat b_{i+1} \hat a_{i+1}) + (\frac{1}{QV}-\frac{1}{YR})
(\hat a^{\dagger}_i \hat b^{\dagger}_i\hat b^{\dagger}_{i+1} \hat d^{\dagger}_{i+1}
\nonumber\\
&+&
\hat a^{\dagger}_i \hat b^{\dagger}_i \hat d_{i+1} \hat b_{i+1} +
\hat b_i \hat a_i \hat b^{\dagger}_{i+1} \hat d^{\dagger}_{i+1} +
\hat b_i \hat a_i \hat d_{i+1} \hat b_{i+1} +
\hat b^{\dagger}_i \hat d^{\dagger}_i\hat a^{\dagger}_{i+1} \hat b^{\dagger}_{i+1}+
\hat b^{\dagger}_i \hat d^{\dagger}_i \hat b_{i+1} \hat a_{i+1} +
\hat d_i \hat b_i \hat a^{\dagger}_{i+1} \hat b^{\dagger}_{i+1}
\nonumber\\
&+&
\hat d_i \hat b_i \hat b_{i+1} \hat a_{i+1}) + (\frac{1}{QW}-\frac{1}{ZR})
(\hat a^{\dagger}_i \hat b^{\dagger}_i\hat e^{\dagger}_{i+1} \hat d^{\dagger}_{i+1} +
\hat a^{\dagger}_i \hat b^{\dagger}_i \hat d_{i+1} \hat e_{i+1} +
\hat b_i \hat a_i \hat e^{\dagger}_{i+1} \hat d^{\dagger}_{i+1} +
\hat b_i \hat a_i \hat d_{i+1} \hat e_{i+1}
\nonumber\\
&+&
\hat e^{\dagger}_i \hat d^{\dagger}_i\hat a^{\dagger}_{i+1} \hat b^{\dagger}_{i+1}+
\hat e^{\dagger}_i \hat d^{\dagger}_i \hat b_{i+1} \hat a_{i+1} +
\hat d_i \hat e_i \hat a^{\dagger}_{i+1} \hat b^{\dagger}_{i+1} +
\hat d_i \hat e_i \hat b_{i+1} \hat a_{i+1}) -(\frac{1}{XV}+\frac{1}{UY})
(\hat a^{\dagger}_i \hat b^{\dagger}_i\hat a_{i+1} \hat b^{\dagger}_{i+1}
\nonumber\\
&+&
\hat a^{\dagger}_i \hat b^{\dagger}_i \hat b_{i+1} \hat a^{\dagger}_{i+1} +
\hat b_i \hat a_i \hat a_{i+1} \hat b^{\dagger}_{i+1} +
\hat b_i \hat a_i \hat b_{i+1} \hat a^{\dagger}_{i+1} +
\hat a_i \hat b^{\dagger}_i\hat a^{\dagger}_{i+1} \hat b^{\dagger}_{i+1}+
\hat a_i \hat b^{\dagger}_i \hat b_{i+1} \hat a_{i+1} +
\hat b_i \hat a^{\dagger}_i \hat a^{\dagger}_{i+1} \hat b^{\dagger}_{i+1}
\nonumber\\
&+&
\hat b_i \hat a^{\dagger}_i \hat b_{i+1} \hat a_{i+1}) -(\frac{1}{XW}+\frac{1}{UZ})
(\hat a^{\dagger}_i \hat b^{\dagger}_i\hat a_{i+1} \hat e^{\dagger}_{i+1} +
\hat a^{\dagger}_i \hat b^{\dagger}_i \hat e_{i+1} \hat a^{\dagger}_{i+1} +
\hat b_i \hat a_i \hat a_{i+1} \hat e^{\dagger}_{i+1} +
\hat b_i \hat a_i \hat e_{i+1} \hat a^{\dagger}_{i+1}
\nonumber\\
&+&
\hat a_i \hat e^{\dagger}_i\hat a^{\dagger}_{i+1} \hat b^{\dagger}_{i+1}+
\hat a_i \hat e^{\dagger}_i \hat b_{i+1} \hat a_{i+1} +
\hat e_i \hat a^{\dagger}_i \hat a^{\dagger}_{i+1} \hat b^{\dagger}_{i+1}+
\hat e_i \hat a^{\dagger}_i \hat b_{i+1} \hat a_{i+1}) -(\frac{1}{XR}+\frac{1}{UQ})
(\hat a^{\dagger}_i \hat b^{\dagger}_i\hat a_{i+1} \hat d^{\dagger}_{i+1}
\nonumber\\
&+&
\hat a^{\dagger}_i \hat b^{\dagger}_i \hat d_{i+1} \hat a^{\dagger}_{i+1} +
\hat b_i \hat a_i \hat a_{i+1} \hat d^{\dagger}_{i+1} +
\hat b_i \hat a_i \hat d_{i+1} \hat a^{\dagger}_{i+1} +
\hat a_i \hat d^{\dagger}_i\hat a^{\dagger}_{i+1} \hat b^{\dagger}_{i+1}+
\hat a_i \hat d^{\dagger}_i \hat b_{i+1} \hat a_{i+1} +
\hat d_i \hat a^{\dagger}_i \hat a^{\dagger}_{i+1} \hat b^{\dagger}_{i+1}
\nonumber\\
&+&
\hat d_i \hat a^{\dagger}_i \hat b_{i+1} \hat a_{i+1}) -(\frac{1}{YW}+\frac{1}{ZV})
(\hat a^{\dagger}_i \hat b^{\dagger}_i\hat b_{i+1} \hat e^{\dagger}_{i+1} +
\hat a^{\dagger}_i \hat b^{\dagger}_i \hat e_{i+1} \hat b^{\dagger}_{i+1} +
\hat b_i \hat a_i \hat b_{i+1} \hat e^{\dagger}_{i+1} +
\hat b_i \hat a_i \hat e_{i+1} \hat b^{\dagger}_{i+1}
\nonumber\\
&+&
\hat b_i \hat e^{\dagger}_i\hat a^{\dagger}_{i+1} \hat b^{\dagger}_{i+1}+
\hat b_i \hat e^{\dagger}_i \hat b_{i+1} \hat a_{i+1} +
\hat e_i \hat b^{\dagger}_i \hat a^{\dagger}_{i+1} \hat b^{\dagger}_{i+1}+
\hat e_i \hat b^{\dagger}_i \hat b_{i+1} \hat a_{i+1}) -(\frac{1}{YR}+\frac{1}{QV})
(\hat a^{\dagger}_i \hat b^{\dagger}_i\hat b_{i+1} \hat d^{\dagger}_{i+1}
\nonumber\\
&+&
\hat a^{\dagger}_i \hat b^{\dagger}_i \hat d_{i+1} \hat b^{\dagger}_{i+1} +
\hat b_i \hat a_i \hat b_{i+1} \hat d^{\dagger}_{i+1} +
\hat b_i \hat a_i \hat d_{i+1} \hat b^{\dagger}_{i+1} +
\hat b_i \hat d^{\dagger}_i\hat a^{\dagger}_{i+1} \hat b^{\dagger}_{i+1}+
\hat b_i \hat d^{\dagger}_i \hat b_{i+1} \hat a_{i+1} +
\hat d_i \hat b^{\dagger}_i \hat a^{\dagger}_{i+1} \hat b^{\dagger}_{i+1}
\nonumber\\
&+&
\hat d_i \hat b^{\dagger}_i \hat b_{i+1} \hat a_{i+1}) -(\frac{1}{ZR}+\frac{1}{QW})
(\hat a^{\dagger}_i \hat b^{\dagger}_i\hat e_{i+1} \hat d^{\dagger}_{i+1} +
\hat a^{\dagger}_i \hat b^{\dagger}_i \hat d_{i+1} \hat e^{\dagger}_{i+1} +
\hat b_i \hat a_i \hat e_{i+1} \hat d^{\dagger}_{i+1} +
\hat b_i \hat a_i \hat d_{i+1} \hat e^{\dagger}_{i+1}
\nonumber\\%
&+&
\hat e_i \hat d^{\dagger}_i\hat a^{\dagger}_{i+1} \hat b^{\dagger}_{i+1}+
\hat e_i \hat d^{\dagger}_i \hat b_{i+1} \hat a_{i+1} +
\hat d_i \hat e^{\dagger}_i \hat a^{\dagger}_{i+1} \hat b^{\dagger}_{i+1} +
\hat d_i \hat e^{\dagger}_i \hat b_{i+1} \hat a_{i+1}) ] \}. 
\label{EQU14}  
\end{eqnarray}
\begin{eqnarray}
\hat P_{5,i} &=& \frac{1}{4} \{ (\frac{1}{UZ}-\frac{1}{XW}) [ (\frac{1}{QU}-
\frac{1}{XR})(\hat a^{\dagger}_i \hat e^{\dagger}_i \hat a^{\dagger}_{i+1}
\hat d^{\dagger}_{i+1} + \hat a^{\dagger}_i \hat e^{\dagger}_i \hat d_{i+1}\hat a_{i+1}
+ \hat e_i \hat a_i \hat a^{\dagger}_{i+1} \hat d^{\dagger}_{i+1}
+ \hat e_i \hat a_i \hat d_{i+1} \hat a_{i+1} 
\nonumber\\
&+&
\hat a^{\dagger}_i \hat d^{\dagger}_i\hat a^{\dagger}_{i+1} \hat e^{\dagger}_{i+1} +
\hat a^{\dagger}_i \hat d^{\dagger}_i\hat e_{i+1} \hat a_{i+1} +
\hat d_i \hat a_i \hat a^{\dagger}_{i+1}\hat e^{\dagger}_{i+1} +
\hat d_i \hat a_i \hat e_{i+1} \hat a_{i+1}) +
(\frac{1}{ZV}-\frac{1}{YW})(\hat a^{\dagger}_i\hat e^{\dagger}_i
\hat b^{\dagger}_{i+1}\hat e^{\dagger}_{i+1}
\nonumber\\
&+&
\hat a^{\dagger}_i \hat e^{\dagger}_i \hat e_{i+1}\hat b_{i+1} +
\hat e_i \hat a_i \hat b^{\dagger}_{i+1} \hat e^{\dagger}_{i+1} +
\hat e_i \hat a_i \hat e_{i+1} \hat b_{i+1} + %
\hat b^{\dagger}_i \hat e^{\dagger}_i\hat a^{\dagger}_{i+1} \hat e^{\dagger}_{i+1} +
\hat b^{\dagger}_i \hat e^{\dagger}_i\hat e_{i+1} \hat a_{i+1} +
\hat e_i \hat b_i \hat a^{\dagger}_{i+1}\hat e^{\dagger}_{i+1}
\nonumber\\
&+&
\hat e_i \hat b_i \hat e_{i+1} \hat a_{i+1}) + (\frac{1}{QV}-\frac{1}{YR})
(\hat a^{\dagger}_i\hat e^{\dagger}_i\hat b^{\dagger}_{i+1}\hat d^{\dagger}_{i+1} +
\hat a^{\dagger}_i \hat e^{\dagger}_i \hat d_{i+1}\hat b_{i+1} +
\hat e_i \hat a_i \hat b^{\dagger}_{i+1} \hat d^{\dagger}_{i+1} +
\hat e_i \hat a_i \hat d_{i+1} \hat b_{i+1}
\nonumber\\%
&+& 
\hat b^{\dagger}_i \hat d^{\dagger}_i\hat a^{\dagger}_{i+1} \hat e^{\dagger}_{i+1} +
\hat b^{\dagger}_i \hat d^{\dagger}_i\hat e_{i+1} \hat a_{i+1} +
\hat d_i \hat b_i \hat a^{\dagger}_{i+1}\hat e^{\dagger}_{i+1} +
\hat d_i \hat b_i \hat e_{i+1} \hat a_{i+1}) + (\frac{1}{QW}-\frac{1}{ZR})
(\hat a^{\dagger}_i\hat e^{\dagger}_i\hat e^{\dagger}_{i+1}\hat d^{\dagger}_{i+1}
\nonumber\\
&+&
\hat a^{\dagger}_i \hat e^{\dagger}_i \hat d_{i+1}\hat e_{i+1} +
\hat e_i \hat a_i \hat e^{\dagger}_{i+1} \hat d^{\dagger}_{i+1} +
\hat e_i \hat a_i \hat d_{i+1} \hat e_{i+1} + 
\hat e^{\dagger}_i \hat d^{\dagger}_i\hat a^{\dagger}_{i+1} \hat e^{\dagger}_{i+1} +
\hat e^{\dagger}_i \hat d^{\dagger}_i\hat e_{i+1} \hat a_{i+1} +
\hat d_i \hat e_i \hat a^{\dagger}_{i+1}\hat e^{\dagger}_{i+1}
\nonumber\\
&+&
\hat d_i \hat e_i \hat e_{i+1} \hat a_{i+1}) -(\frac{1}{XV}+\frac{1}{UY})
(\hat a^{\dagger}_i \hat e^{\dagger}_i\hat a_{i+1} \hat b^{\dagger}_{i+1} +
\hat a^{\dagger}_i \hat e^{\dagger}_i \hat b_{i+1} \hat a^{\dagger}_{i+1} +
\hat e_i \hat a_i \hat a_{i+1} \hat b^{\dagger}_{i+1} +
\hat e_i \hat a_i \hat b_{i+1} \hat a^{\dagger}_{i+1}
\nonumber\\
&+&
\hat a_i \hat b^{\dagger}_i\hat a^{\dagger}_{i+1} \hat e^{\dagger}_{i+1}+
\hat a_i \hat b^{\dagger}_i \hat e_{i+1} \hat a_{i+1} +
\hat b_i \hat a^{\dagger}_i \hat a^{\dagger}_{i+1} \hat e^{\dagger}_{i+1} +
\hat b_i \hat a^{\dagger}_i \hat e_{i+1} \hat a_{i+1}) -(\frac{1}{XW}+\frac{1}{UZ})
(\hat a^{\dagger}_i \hat e^{\dagger}_i\hat a_{i+1} \hat e^{\dagger}_{i+1}
\nonumber\\%
&+&
\hat a^{\dagger}_i \hat e^{\dagger}_i \hat e_{i+1} \hat a^{\dagger}_{i+1} +
\hat e_i \hat a_i \hat a_{i+1} \hat e^{\dagger}_{i+1} +
\hat e_i \hat a_i \hat e_{i+1} \hat a^{\dagger}_{i+1} +
\hat a_i \hat e^{\dagger}_i\hat a^{\dagger}_{i+1} \hat e^{\dagger}_{i+1} +
\hat a_i \hat e^{\dagger}_i \hat e_{i+1} \hat a_{i+1} +
\hat e_i \hat a^{\dagger}_i \hat a^{\dagger}_{i+1} \hat e^{\dagger}_{i+1} 
\nonumber\\
 &+&
\hat e_i \hat a^{\dagger}_i \hat e_{i+1} \hat a_{i+1}) -(\frac{1}{XR}+\frac{1}{UQ})
(\hat a^{\dagger}_i \hat e^{\dagger}_i\hat a_{i+1} \hat d^{\dagger}_{i+1} +
\hat a^{\dagger}_i \hat e^{\dagger}_i \hat d_{i+1} \hat a^{\dagger}_{i+1} +
\hat e_i \hat a_i \hat a_{i+1} \hat d^{\dagger}_{i+1} +
\hat e_i \hat a_i \hat d_{i+1} \hat a^{\dagger}_{i+1}
\nonumber\\
&+&
\hat a_i \hat d^{\dagger}_i\hat a^{\dagger}_{i+1} \hat e^{\dagger}_{i+1}+
\hat a_i \hat d^{\dagger}_i \hat e_{i+1} \hat a_{i+1} +
\hat d_i \hat a^{\dagger}_i \hat a^{\dagger}_{i+1} \hat e^{\dagger}_{i+1} +
\hat d_i \hat a^{\dagger}_i \hat e_{i+1} \hat a_{i+1}) -(\frac{1}{YW}+\frac{1}{ZV})
(\hat a^{\dagger}_i \hat e^{\dagger}_i\hat b_{i+1} \hat e^{\dagger}_{i+1}
\nonumber\\
&+&
\hat a^{\dagger}_i \hat e^{\dagger}_i \hat e_{i+1} \hat b^{\dagger}_{i+1} +
\hat e_i \hat a_i \hat b_{i+1} \hat e^{\dagger}_{i+1} +
\hat e_i \hat a_i \hat e_{i+1} \hat b^{\dagger}_{i+1} +
\hat b_i \hat e^{\dagger}_i\hat a^{\dagger}_{i+1} \hat e^{\dagger}_{i+1}+
\hat b_i \hat e^{\dagger}_i \hat e_{i+1} \hat a_{i+1} +
\hat e_i \hat b^{\dagger}_i \hat a^{\dagger}_{i+1} \hat e^{\dagger}_{i+1}
\nonumber\\%
&+&
\hat e_i \hat b^{\dagger}_i \hat e_{i+1} \hat a_{i+1}) -(\frac{1}{YR}+\frac{1}{QV})
(\hat a^{\dagger}_i \hat e^{\dagger}_i\hat b_{i+1} \hat d^{\dagger}_{i+1} +
\hat a^{\dagger}_i \hat e^{\dagger}_i \hat d_{i+1} \hat b^{\dagger}_{i+1} +
\hat e_i \hat a_i \hat b_{i+1} \hat d^{\dagger}_{i+1} +
\hat e_i \hat a_i \hat d_{i+1} \hat b^{\dagger}_{i+1}
\nonumber\\
&+&
\hat b_i \hat d^{\dagger}_i\hat a^{\dagger}_{i+1} \hat e^{\dagger}_{ i+1}+
\hat b_i \hat d^{\dagger}_i \hat e_{i+1} \hat a_{i+1} +
\hat d_i \hat b^{\dagger}_i \hat a^{\dagger}_{i+1} \hat e^{\dagger}_{i+1} +
\hat d_i \hat b^{\dagger}_i \hat e_{i+1} \hat a_{i+1}) -(\frac{1}{ZR}+\frac{1}{QW})
(\hat a^{\dagger}_i \hat e^{\dagger}_i\hat e_{i+1} \hat d^{\dagger}_{i+1}
\nonumber\\
&+&
\hat a^{\dagger}_i \hat e^{\dagger}_i \hat d_{i+1} \hat e^{\dagger}_{i+1} +
\hat e_i \hat a_i \hat e_{i+1} \hat d^{\dagger}_{i+1} +
\hat e_i \hat a_i \hat d_{i+1} \hat e^{\dagger}_{i+1} +
\hat e_i \hat d^{\dagger}_i\hat a^{\dagger}_{i+1} \hat e^{\dagger}_{i+1}+
\hat e_i \hat d^{\dagger}_i \hat e_{i+1} \hat a_{i+1} +
\hat d_i \hat e^{\dagger}_i \hat a^{\dagger}_{i+1} \hat e^{\dagger}_{i+1}
\nonumber\\
&+&
\hat d_i \hat e^{\dagger}_i \hat e_{i+1} \hat a_{i+1}) ] \}. 
\label{EQU15}  
\end{eqnarray}
\begin{eqnarray}
\hat P_{6,i} &=& \frac{1}{4} \{ (\frac{1}{QU}-\frac{1}{XR}) [ (\frac{1}{ZV}-
\frac{1}{YW})(\hat a^{\dagger}_i \hat d^{\dagger}_i \hat b^{\dagger}_{i+1}
\hat e^{\dagger}_{i+1} + \hat a^{\dagger}_i \hat d^{\dagger}_i \hat e_{i+1}\hat b_{i+1}
+ \hat d_i \hat a_i \hat b^{\dagger}_{i+1} \hat e^{\dagger}_{i+1}
+ \hat d_i \hat a_i \hat e_{i+1} \hat b_{i+1} 
\nonumber\\
&+&
\hat b^{\dagger}_i \hat e^{\dagger}_i\hat a^{\dagger}_{i+1} \hat d^{\dagger}_{i+1} +
\hat b^{\dagger}_i \hat e^{\dagger}_i\hat d_{i+1} \hat a_{i+1} +
\hat e_i \hat b_i \hat a^{\dagger}_{i+1}\hat d^{\dagger}_{i+1} +
\hat e_i \hat b_i \hat d_{i+1} \hat a_{i+1}) +
(\frac{1}{QV}-\frac{1}{YR})(\hat a^{\dagger}_i\hat d^{\dagger}_i
\hat b^{\dagger}_{i+1}\hat d^{\dagger}_{i+1}
\nonumber\\
&+&
\hat a^{\dagger}_i \hat d^{\dagger}_i \hat d_{i+1}\hat b_{i+1} +
\hat d_i \hat a_i \hat b^{\dagger}_{i+1} \hat d^{\dagger}_{i+1} +
\hat d_i \hat a_i \hat d_{i+1} \hat b_{i+1} + 
\hat b^{\dagger}_i \hat d^{\dagger}_i\hat a^{\dagger}_{i+1} \hat d^{\dagger}_{i+1} +
\hat b^{\dagger}_i \hat d^{\dagger}_i\hat d_{i+1} \hat a_{i+1} +
\hat d_i \hat b_i \hat a^{\dagger}_{i+1}\hat d^{\dagger}_{i+1}
\nonumber\\%
&+&
\hat d_i \hat b_i \hat d^{\dagger}_{i+1}\hat a^{\dagger}_{i+1}) +
(\frac{1}{QW}-\frac{1}{ZR})(\hat a^{\dagger}_i \hat d^{\dagger}_i
\hat e^{\dagger}_{i+1}\hat d^{\dagger}_{i+1} +
\hat a^{\dagger}_i \hat d^{\dagger}_i \hat d_{i+1}\hat e_{i+1}
+ \hat d_i \hat a_i \hat e^{\dagger}_{i+1} \hat d^{\dagger}_{i+1}
+ \hat d_i \hat a_i \hat d_{i+1} \hat e_{i+1}
\nonumber\\%
&+&
\hat e^{\dagger}_i \hat d^{\dagger}_i\hat a^{\dagger}_{i+1} \hat d^{\dagger}_{i+1} +
\hat e^{\dagger}_i \hat d^{\dagger}_i\hat d_{i+1} \hat a_{i+1} +
\hat d_i \hat e_i \hat a^{\dagger}_{i+1}\hat d^{\dagger}_{i+1} +
\hat d_i \hat e_i \hat d_{i+1} \hat a_{i+1}) 
-(\frac{1}{XV}+\frac{1}{UY})
(\hat a^{\dagger}_i \hat d^{\dagger}_i\hat a_{i+1} \hat b^{\dagger}_{i+1}
\nonumber\\
&+&
\hat a^{\dagger}_i \hat d^{\dagger}_i \hat b_{i+1} \hat a^{\dagger}_{i+1} +
\hat d_i \hat a_i \hat a_{i+1} \hat b^{\dagger}_{i+1} +
\hat d_i \hat a_i \hat b_{i+1} \hat a^{\dagger}_{i+1} +
\hat a_i \hat b^{\dagger}_i\hat a^{\dagger}_{i+1} \hat d^{\dagger}_{i+1}+
\hat a_i \hat b^{\dagger}_i \hat d_{i+1} \hat a_{i+1} +
\hat b_i \hat a^{\dagger}_i \hat a^{\dagger}_{i+1} \hat d^{\dagger}_{i+1}
\nonumber\\
&+&
\hat b_i \hat a^{\dagger}_i \hat d_{i+1} \hat a_{i+1}) -(\frac{1}{XW}+\frac{1}{UZ})
(\hat a^{\dagger}_i \hat d^{\dagger}_i\hat a_{i+1} \hat e^{\dagger}_{i+1} +
\hat a^{\dagger}_i \hat d^{\dagger}_i \hat e_{i+1} \hat a^{\dagger}_{i+1} +
\hat d_i \hat a_i \hat a_{i+1} \hat e^{\dagger}_{i+1} +
\hat d_i \hat a_i \hat e_{i+1} \hat a^{\dagger}_{i+1}
\nonumber\\
&+&
\hat a_i \hat e^{\dagger}_i\hat a^{\dagger}_{i+1} \hat d^{\dagger}_{i+1}+
\hat a_i \hat e^{\dagger}_i \hat d_{i+1} \hat a_{i+1} +
\hat e_i \hat a^{\dagger}_i \hat a^{\dagger}_{i+1} \hat d^{\dagger}_{i+1}+
\hat e_i \hat a^{\dagger}_i \hat d_{i+1} \hat a_{i+1}) -(\frac{1}{XR}+\frac{1}{UQ})
(\hat a^{\dagger}_i \hat d^{\dagger}_i\hat a_{i+1} \hat d^{\dagger}_{i+1}
\nonumber\\%
&+&
\hat a^{\dagger}_i \hat d^{\dagger}_i \hat d_{i+1} \hat a^{\dagger}_{i+1} +
\hat d_i \hat a_i \hat a_{i+1} \hat d^{\dagger}_{i+1} +
\hat d_i \hat a_i \hat d_{i+1} \hat a^{\dagger}_{i+1} +
\hat a_i \hat d^{\dagger}_i\hat a^{\dagger}_{i+1} \hat d^{\dagger}_{i+1}+
\hat a_i \hat d^{\dagger}_i \hat d_{i+1} \hat a_{i+1} +
\hat d_i \hat a^{\dagger}_i \hat a^{\dagger}_{i+1} \hat d^{\dagger}_{i+1}
\nonumber\\
&+&
\hat d_i \hat a^{\dagger}_i \hat d_{i+1} \hat a_{i+1}) -(\frac{1}{YW}+\frac{1}{ZV})
(\hat a^{\dagger}_i \hat d^{\dagger}_i\hat b_{i+1} \hat e^{\dagger}_{i+1} +
\hat a^{\dagger}_i \hat d^{\dagger}_i \hat e_{i+1} \hat b^{\dagger}_{i+1} +
\hat d_i \hat a_i \hat b_{i+1} \hat e^{\dagger}_{i+1} +
\hat d_i \hat a_i \hat e_{i+1} \hat b^{\dagger}_{i+1}
\nonumber\\
&+&
\hat b_i \hat e^{\dagger}_i\hat a^{\dagger}_{i+1} \hat d^{\dagger}_{i+1}+
\hat b_i \hat e^{\dagger}_i \hat d_{i+1} \hat a_{i+1} +
\hat e_i \hat b^{\dagger}_i \hat a^{\dagger}_{i+1} \hat d^{\dagger}_{i+1}+
\hat e_i \hat b^{\dagger}_i \hat d_{i+1} \hat a_{i+1}) -(\frac{1}{YR}+\frac{1}{QV})
(\hat a^{\dagger}_i \hat d^{\dagger}_i\hat b_{i+1} \hat d^{\dagger}_{i+1}
\nonumber\\
&+&
\hat a^{\dagger}_i \hat d^{\dagger}_i \hat d_{i+1} \hat b^{\dagger}_{i+1} +
\hat d_i \hat a_i \hat b_{i+1} \hat d^{\dagger}_{i+1} +
\hat d_i \hat a_i \hat d_{i+1} \hat b^{\dagger}_{i+1} +
\hat b_i \hat d^{\dagger}_i\hat a^{\dagger}_{i+1} \hat d^{\dagger}_{i+1}+
\hat b_i \hat d^{\dagger}_i \hat d_{i+1} \hat a_{i+1} +
\hat d_i \hat b^{\dagger}_i \hat a^{\dagger}_{i+1} \hat d^{\dagger}_{i+1}
\nonumber\\
&+&
\hat d_i \hat b^{\dagger}_i \hat d_{i+1} \hat a_{i+1}) -(\frac{1}{ZR}+\frac{1}{QW})
(\hat a^{\dagger}_i \hat d^{\dagger}_i\hat e_{i+1} \hat d^{\dagger}_{i+1} +
\hat a^{\dagger}_i \hat d^{\dagger}_i \hat d_{i+1} \hat e^{\dagger}_{i+1} +
\hat d_i \hat a_i \hat e_{i+1} \hat d^{\dagger}_{i+1} +
\hat d_i \hat a_i \hat d_{i+1} \hat e^{\dagger}_{i+1}
\nonumber\\
&+&
\hat e_i \hat d^{\dagger}_i\hat a^{\dagger}_{i+1} \hat d^{\dagger}_{i+1}+
\hat e_i \hat d^{\dagger}_i \hat d_{i+1} \hat a_{i+1} +
\hat d_i \hat e^{\dagger}_i \hat a^{\dagger}_{i+1} \hat d^{\dagger}_{i+1} +
\hat d_i \hat e^{\dagger}_i \hat d_{i+1} \hat a_{i+1}) ] \}. 
\label{EQU16}
\end{eqnarray}
\begin{eqnarray}
\hat P_{7,i} &=& \frac{1}{4} \{ (\frac{1}{ZV}-\frac{1}{YW}) [ (\frac{1}{QV}-
\frac{1}{YR})(\hat b^{\dagger}_i \hat e^{\dagger}_i \hat b^{\dagger}_{i+1}
\hat d^{\dagger}_{i+1} + \hat b^{\dagger}_i \hat e^{\dagger}_i \hat d_{i+1}\hat b_{i+1}
+ \hat e_i \hat b_i \hat b^{\dagger}_{i+1} \hat d^{\dagger}_{i+1}
+ \hat e_i \hat b_i \hat d_{i+1} \hat b_{i+1} 
\nonumber\\
&+&
\hat b^{\dagger}_i \hat d^{\dagger}_i\hat b^{\dagger}_{i+1} \hat e^{\dagger}_{i+1} +
\hat b^{\dagger}_i \hat d^{\dagger}_i\hat e_{i+1} \hat b_{i+1} +
\hat d_i \hat b_i \hat b^{\dagger}_{i+1}\hat e^{\dagger}_{i+1} +
\hat d_i \hat b_i \hat e_{i+1} \hat b_{i+1}) +
(\frac{1}{QW}-\frac{1}{ZR})(\hat b^{\dagger}_i\hat e^{\dagger}_i
\hat e^{\dagger}_{i+1}\hat d^{\dagger}_{i+1}
\nonumber\\
&+&
\hat b^{\dagger}_i \hat e^{\dagger}_i \hat d_{i+1}\hat e_{i+1} +
\hat e_i \hat b_i \hat e^{\dagger}_{i+1} \hat d^{\dagger}_{i+1} +
\hat e_i \hat b_i \hat d_{i+1} \hat e_{i+1} + 
\hat e^{\dagger}_i \hat d^{\dagger}_i\hat b^{\dagger}_{i+1} \hat e^{\dagger}_{i+1} +
\hat e^{\dagger}_i \hat d^{\dagger}_i\hat e_{i+1} \hat b_{i+1} +
\hat d_i \hat e_i \hat b^{\dagger}_{i+1}\hat e^{\dagger}_{i+1}
\nonumber\\%
&+&
\hat d_i \hat e_i \hat e_{i+1}\hat b_{i+1}) 
-(\frac{1}{XV}+\frac{1}{UY})
(\hat b^{\dagger}_i \hat e^{\dagger}_i\hat a_{i+1} \hat b^{\dagger}_{i+1} +
\hat b^{\dagger}_i \hat e^{\dagger}_i \hat b_{i+1} \hat a^{\dagger}_{i+1} +
\hat e_i \hat b_i \hat a_{i+1} \hat b^{\dagger}_{i+1} +
\hat e_i \hat b_i \hat b_{i+1} \hat a^{\dagger}_{i+1}
\nonumber\\
&+&
\hat a_i \hat b^{\dagger}_i\hat b^{\dagger}_{i+1} \hat e^{\dagger}_{i+1}+
\hat a_i \hat b^{\dagger}_i \hat e_{i+1} \hat b_{i+1} +
\hat b_i \hat a^{\dagger}_i \hat b^{\dagger}_{i+1} \hat e^{\dagger}_{i+1} +
\hat b_i \hat a^{\dagger}_i \hat e_{i+1} \hat b_{i+1}) -(\frac{1}{XW}+\frac{1}{UZ})
(\hat b^{\dagger}_i \hat e^{\dagger}_i\hat a_{i+1} \hat e^{\dagger}_{i+1}
\nonumber\\
&+&
\hat b^{\dagger}_i \hat e^{\dagger}_i \hat e_{i+1} \hat a^{\dagger}_{i+1} +
\hat e_i \hat b_i \hat a_{i+1} \hat e^{\dagger}_{i+1} +
\hat e_i \hat b_i \hat e_{i+1} \hat a^{\dagger}_{i+1} +
\hat a_i \hat e^{\dagger}_i\hat b^{\dagger}_{i+1} \hat e^{\dagger}_{i+1}+
\hat a_i \hat e^{\dagger}_i \hat e_{i+1} \hat b_{i+1} +
\hat e_i \hat a^{\dagger}_i \hat b^{\dagger}_{i+1} \hat e^{\dagger}_{i+1}
\nonumber\\
&+& 
\hat e_i \hat a^{\dagger}_i \hat e_{i+1} \hat b_{i+1}) -(\frac{1}{XR}+\frac{1}{UQ})
(\hat b^{\dagger}_i \hat e^{\dagger}_i\hat a_{i+1} \hat d^{\dagger}_{i+1} +
\hat b^{\dagger}_i \hat e^{\dagger}_i \hat d_{i+1} \hat a^{\dagger}_{i+1} +
\hat e_i \hat b_i \hat a_{i+1} \hat d^{\dagger}_{i+1} +
\hat e_i \hat b_i \hat d_{i+1} \hat a^{\dagger}_{i+1}
\nonumber\\
&+&
\hat a_i \hat d^{\dagger}_i\hat b^{\dagger}_{i+1} \hat e^{\dagger}_{i+1}+
\hat a_i \hat d^{\dagger}_i \hat e_{i+1} \hat b_{i+1} +
\hat d_i \hat a^{\dagger}_i \hat b^{\dagger}_{i+1} \hat e^{\dagger}_{i+1} +
\hat d_i \hat a^{\dagger}_i \hat e_{i+1} \hat b_{i+1}) -(\frac{1}{YW}+\frac{1}{ZV})
(\hat b^{\dagger}_i \hat e^{\dagger}_i\hat b_{i+1} \hat e^{\dagger}_{i+1}
\nonumber\\
&+&
\hat b^{\dagger}_i \hat e^{\dagger}_i \hat e_{i+1} \hat b^{\dagger}_{i+1} +
\hat e_i \hat b_i \hat b_{i+1} \hat e^{\dagger}_{i+1} +
\hat e_i \hat b_i \hat e_{i+1} \hat b^{\dagger}_{i+1} +
\hat b_i \hat e^{\dagger}_i\hat b^{\dagger}_{i+1} \hat e^{\dagger}_{i+1}+
\hat b_i \hat e^{\dagger}_i \hat e_{i+1} \hat b_{i+1} +
\hat e_i \hat b^{\dagger}_i \hat b^{\dagger}_{i+1} \hat e^{\dagger}_{i+1}
\nonumber\\
&+&
\hat e_i \hat b^{\dagger}_i \hat e_{i+1} \hat b_{i+1}) -(\frac{1}{YR}+\frac{1}{QV})
(\hat b^{\dagger}_i \hat e^{\dagger}_i\hat b_{i+1} \hat d^{\dagger}_{i+1} +
\hat b^{\dagger}_i \hat e^{\dagger}_i \hat d_{i+1} \hat b^{\dagger}_{i+1} +
\hat e_i \hat b_i \hat b_{i+1} \hat d^{\dagger}_{i+1} +
\hat e_i \hat b_i \hat d_{i+1} \hat b^{\dagger}_{i+1}
\nonumber\\
&+&
\hat b_i \hat d^{\dagger}_i\hat b^{\dagger}_{i+1} \hat e^{\dagger}_{i+1}+
\hat b_i \hat d^{\dagger}_i \hat e_{i+1} \hat b_{i+1} +
\hat d_i \hat b^{\dagger}_i \hat b^{\dagger}_{i+1} \hat e^{\dagger}_{i+1} +
\hat d_i \hat b^{\dagger}_i \hat e_{i+1} \hat b_{i+1}) -(\frac{1}{ZR}+\frac{1}{QW})
(\hat b^{\dagger}_i \hat e^{\dagger}_i\hat e_{i+1} \hat d^{\dagger}_{i+1}
\nonumber\\
&+&
\hat b^{\dagger}_i \hat e^{\dagger}_i \hat d_{i+1} \hat e^{\dagger}_{i+1} +
\hat e_i \hat b_i \hat e_{i+1} \hat d^{\dagger}_{i+1} +
\hat e_i \hat b_i \hat d_{i+1} \hat e^{\dagger}_{i+1} +
\hat e_i \hat d^{\dagger}_i\hat b^{\dagger}_{i+1} \hat e^{\dagger}_{i+1}+
\hat e_i \hat d^{\dagger}_i \hat e_{i+1} \hat b_{i+1} +
\hat d_i \hat e^{\dagger}_i \hat b^{\dagger}_{i+1} \hat e^{\dagger}_{i+1}
\nonumber\\
&+&
\hat d_i \hat e^{\dagger}_i \hat e_{i+1} \hat b_{i+1}) ] \}. 
\label{EQU17}  
\end{eqnarray}
\begin{eqnarray}
\hat P_{8,i} &=& \frac{1}{4} \{ (\frac{1}{QV}-\frac{1}{YR}) [ (\frac{1}{QW}-
\frac{1}{ZR})(\hat b^{\dagger}_i \hat d^{\dagger}_i \hat e^{\dagger}_{i+1}
\hat d^{\dagger}_{i+1} + \hat b^{\dagger}_i \hat d^{\dagger}_i \hat d_{i+1}\hat e_{i+1}
+ \hat d_i \hat b_i \hat e^{\dagger}_{i+1} \hat d^{\dagger}_{i+1}
+ \hat d_i \hat b_i \hat d_{i+1} \hat e_{i+1} 
\nonumber\\
&+&
\hat e^{\dagger}_i \hat d^{\dagger}_i\hat b^{\dagger}_{i+1} \hat d^{\dagger}_{i+1} +
\hat e^{\dagger}_i \hat d^{\dagger}_i\hat d_{i+1} \hat b_{i+1} +
\hat d_i \hat e_i \hat b^{\dagger}_{i+1}\hat d^{\dagger}_{i+1} +
\hat d_i \hat e_i \hat d_{i+1} \hat b_{i+1}) -
(\frac{1}{XV}+\frac{1}{UY})
(\hat b^{\dagger}_i \hat d^{\dagger}_i\hat a_{i+1} \hat b^{\dagger}_{i+1}
\nonumber\\
&+&
\hat b^{\dagger}_i \hat d^{\dagger}_i \hat b_{i+1} \hat a^{\dagger}_{i+1} +
\hat d_i \hat b_i \hat a_{i+1} \hat b^{\dagger}_{i+1} +
\hat d_i \hat b_i \hat b_{i+1} \hat a^{\dagger}_{i+1} +
\hat a_i \hat b^{\dagger}_i\hat b^{\dagger}_{i+1} \hat d^{\dagger}_{i+1}+
\hat a_i \hat b^{\dagger}_i \hat d_{i+1} \hat b_{i+1} +
\hat b_i \hat a^{\dagger}_i \hat b^{\dagger}_{i+1} \hat d^{\dagger}_{i+1}
\nonumber\\
&+&
\hat b_i \hat a^{\dagger}_i \hat d_{i+1} \hat b_{i+1}) -(\frac{1}{XW}+\frac{1}{UZ})
(\hat b^{\dagger}_i \hat d^{\dagger}_i\hat a_{i+1} \hat e^{\dagger}_{i+1} +
\hat b^{\dagger}_i \hat d^{\dagger}_i \hat e_{i+1} \hat a^{\dagger}_{i+1} +
\hat d_i \hat b_i \hat a_{i+1} \hat e^{\dagger}_{i+1} +
\hat d_i \hat b_i \hat e_{i+1} \hat a^{\dagger}_{i+1}
\nonumber\\%
&+&
\hat a_i \hat e^{\dagger}_i\hat b^{\dagger}_{i+1} \hat d^{\dagger}_{i+1}+
\hat a_i \hat e^{\dagger}_i \hat d_{i+1} \hat b_{i+1} +
\hat e_i \hat a^{\dagger}_i \hat b^{\dagger}_{i+1} \hat d^{\dagger}_{i+1}+
\hat e_i \hat a^{\dagger}_i \hat d_{i+1} \hat b_{i+1}) -(\frac{1}{XR}+\frac{1}{UQ})
(\hat b^{\dagger}_i \hat d^{\dagger}_i\hat a_{i+1} \hat d^{\dagger}_{i+1}
\nonumber\\
&+&
\hat b^{\dagger}_i \hat d^{\dagger}_i \hat d_{i+1} \hat a^{\dagger}_{i+1} +
\hat d_i \hat b_i \hat a_{i+1} \hat d^{\dagger}_{i+1} +
\hat d_i \hat b_i \hat d_{i+1} \hat a^{\dagger}_{i+1} +
\hat a_i \hat d^{\dagger}_i\hat b^{\dagger}_{i+1} \hat d^{\dagger}_{i+1}+
\hat a_i \hat d^{\dagger}_i \hat d_{i+1} \hat b_{i+1} +
\hat d_i \hat a^{\dagger}_i \hat b^{\dagger}_{i+1} \hat d^{\dagger}_{i+1}
\nonumber\\
&+&
\hat d_i \hat a^{\dagger}_i \hat d_{i+1} \hat b_{i+1}) -(\frac{1}{YW}+\frac{1}{ZV})
(\hat b^{\dagger}_i \hat d^{\dagger}_i\hat b_{i+1} \hat e^{\dagger}_{i+1} +
\hat b^{\dagger}_i \hat d^{\dagger}_i \hat e_{i+1} \hat b^{\dagger}_{i+1} +
\hat d_i \hat b_i \hat b_{i+1} \hat e^{\dagger}_{i+1} +
\hat d_i \hat b_i \hat e_{i+1} \hat b^{\dagger}_{i+1}
\nonumber\\
&+&
\hat b_i \hat e^{\dagger}_i\hat b^{\dagger}_{i+1} \hat d^{\dagger}_{i+1}+
\hat b_i \hat e^{\dagger}_i \hat d_{i+1} \hat b_{i+1} +
\hat e_i \hat b^{\dagger}_i \hat b^{\dagger}_{i+1} \hat d^{\dagger}_{i+1}+
\hat e_i \hat b^{\dagger}_i \hat d_{i+1} \hat b_{i+1}) -(\frac{1}{YR}+\frac{1}{QV})
(\hat b^{\dagger}_i \hat d^{\dagger}_i\hat b_{i+1} \hat d^{\dagger}_{i+1}
\nonumber\\
&+&
\hat b^{\dagger}_i \hat d^{\dagger}_i \hat d_{i+1} \hat b^{\dagger}_{i+1} +
\hat d_i \hat b_i \hat b_{i+1} \hat d^{\dagger}_{i+1} +
\hat d_i \hat b_i \hat d_{i+1} \hat b^{\dagger}_{i+1} +
\hat b_i \hat d^{\dagger}_i\hat b^{\dagger}_{i+1} \hat d^{\dagger}_{i+1}+
\hat b_i \hat d^{\dagger}_i \hat d_{i+1} \hat b_{i+1} +
\hat d_i \hat b^{\dagger}_i \hat b^{\dagger}_{i+1} \hat d^{\dagger}_{i+1}
\nonumber\\
&+&
\hat d_i \hat b^{\dagger}_i \hat d_{i+1} \hat b_{i+1}) -(\frac{1}{ZR}+\frac{1}{QW})
(\hat b^{\dagger}_i \hat d^{\dagger}_i\hat e_{i+1} \hat d^{\dagger}_{i+1} +
\hat b^{\dagger}_i \hat d^{\dagger}_i \hat d_{i+1} \hat e^{\dagger}_{i+1} +
\hat d_i \hat b_i \hat e_{i+1} \hat d^{\dagger}_{i+1} +
\hat d_i \hat b_i \hat d_{i+1} \hat e^{\dagger}_{i+1}
\nonumber\\
&+&
\hat e_i \hat d^{\dagger}_i\hat b^{\dagger}_{i+1} \hat d^{\dagger}_{i+1}+
\hat e_i \hat d^{\dagger}_i \hat d_{i+1} \hat b_{i+1} +
\hat d_i \hat e^{\dagger}_i \hat b^{\dagger}_{i+1} \hat d^{\dagger}_{i+1} +
\hat d_i \hat e^{\dagger}_i \hat d_{i+1} \hat b_{i+1}) ] \}. 
\label{EQU18}
\end{eqnarray}
\begin{eqnarray}
\hat P_{9,i} &=& - \frac{1}{4} \{ (\frac{1}{QW}-\frac{1}{ZR}) [ (\frac{1}{XV}+
\frac{1}{UY})(\hat e^{\dagger}_i \hat d^{\dagger}_i\hat a_{i+1} \hat b^{\dagger}_{i+1}
+\hat e^{\dagger}_i \hat d^{\dagger}_i \hat b_{i+1} \hat a^{\dagger}_{i+1} +
\hat d_i \hat e_i \hat a_{i+1} \hat b^{\dagger}_{i+1} +
\hat d_i \hat e_i \hat b_{i+1} \hat a^{\dagger}_{i+1}
\nonumber\\
&+&
\hat a_i \hat b^{\dagger}_i\hat e^{\dagger}_{i+1} \hat d^{\dagger}_{i+1}+
\hat a_i \hat b^{\dagger}_i \hat d_{i+1} \hat e_{i+1} +
\hat b_i \hat a^{\dagger}_i \hat e^{\dagger}_{i+1} \hat d^{\dagger}_{i+1} +
\hat b_i \hat a^{\dagger}_i \hat d_{i+1} \hat e_{i+1}) +(\frac{1}{XW}+\frac{1}{UZ})
(\hat e^{\dagger}_i \hat d^{\dagger}_i\hat a_{i+1} \hat e^{\dagger}_{i+1}
\nonumber\\
&+&
\hat e^{\dagger}_i \hat d^{\dagger}_i \hat e_{i+1} \hat a^{\dagger}_{i+1} +
\hat d_i \hat e_i \hat a_{i+1} \hat e^{\dagger}_{i+1} +
\hat d_i \hat e_i \hat e_{i+1} \hat a^{\dagger}_{i+1} +
\hat a_i \hat e^{\dagger}_i\hat e^{\dagger}_{i+1} \hat d^{\dagger}_{i+1}+
\hat a_i \hat e^{\dagger}_i \hat d_{i+1} \hat e_{i+1} +
\hat e_i \hat a^{\dagger}_i \hat e^{\dagger}_{i+1} \hat d^{\dagger}_{i+1}
\nonumber\\
&+&
\hat e_i \hat a^{\dagger}_i \hat d_{i+1} \hat e_{i+1}) +(\frac{1}{XR}+\frac{1}{UQ})
(\hat e^{\dagger}_i \hat d^{\dagger}_i\hat a_{i+1} \hat d^{\dagger}_{i+1} +
\hat e^{\dagger}_i \hat d^{\dagger}_i \hat d_{i+1} \hat a^{\dagger}_{i+1} +
\hat d_i \hat e_i \hat a_{i+1} \hat d^{\dagger}_{i+1} +
\hat d_i \hat e_i \hat d_{i+1} \hat a^{\dagger}_{i+1}
\nonumber\\%
&+&
\hat a_i \hat d^{\dagger}_i\hat e^{\dagger}_{i+1} \hat d^{\dagger}_{i+1}+
\hat a_i \hat d^{\dagger}_i \hat d_{i+1} \hat e_{i+1} +
\hat d_i \hat a^{\dagger}_i \hat e^{\dagger}_{i+1} \hat d^{\dagger}_{i+1} +
\hat d_i \hat a^{\dagger}_i \hat d_{i+1} \hat e_{i+1}) +
(\frac{1}{YW}+\frac{1}{ZV})
(\hat e^{\dagger}_i \hat d^{\dagger}_i\hat b_{i+1} \hat e^{\dagger}_{i+1}
\nonumber\\
&+&
\hat e^{\dagger}_i \hat d^{\dagger}_i \hat e_{i+1} \hat b^{\dagger}_{i+1} +
\hat d_i \hat e_i \hat b_{i+1} \hat e^{\dagger}_{i+1} +
\hat d_i \hat e_i \hat e_{i+1} \hat b^{\dagger}_{i+1} +
\hat b_i \hat e^{\dagger}_i\hat e^{\dagger}_{i+1} \hat d^{\dagger}_{i+1}+
\hat b_i \hat e^{\dagger}_i \hat d_{i+1} \hat e_{i+1} +
\hat e_i \hat b^{\dagger}_i \hat e^{\dagger}_{i+1} \hat d^{\dagger}_{i+1}
\nonumber\\
&+&
\hat e_i \hat b^{\dagger}_i \hat d_{i+1} \hat e_{i+1}) +(\frac{1}{YR}+\frac{1}{QV})
(\hat e^{\dagger}_i \hat d^{\dagger}_i\hat b_{i+1} \hat d^{\dagger}_{i+1} +
\hat e^{\dagger}_i \hat d^{\dagger}_i \hat d_{i+1} \hat b^{\dagger}_{i+1} +
\hat d_i \hat e_i \hat b_{i+1} \hat d^{\dagger}_{i+1} +
\hat d_i \hat e_i \hat d_{i+1} \hat b^{\dagger}_{i+1}
\nonumber\\
&+&
\hat b_i \hat d^{\dagger}_i\hat e^{\dagger}_{i+1} \hat d^{\dagger}_{i+1}+
\hat b_i \hat d^{\dagger}_i \hat d_{i+1} \hat e_{i+1} +
\hat d_i \hat b^{\dagger}_i \hat e^{\dagger}_{i+1} \hat d^{\dagger}_{i+1} +
\hat d_i \hat b^{\dagger}_i \hat d_{i+1} \hat e_{i+1}) +(\frac{1}{ZR}+\frac{1}{QW})
(\hat e^{\dagger}_i \hat d^{\dagger}_i\hat e_{i+1} \hat d^{\dagger}_{i+1} 
\nonumber\\
&+&
\hat e^{\dagger}_i \hat d^{\dagger}_i \hat d_{i+1} \hat e^{\dagger}_{i+1} +
\hat d_i \hat e_i \hat e_{i+1} \hat d^{\dagger}_{i+1} +
\hat d_i \hat e_i \hat d_{i+1} \hat e^{\dagger}_{i+1} +
\hat e_i \hat d^{\dagger}_i\hat e^{\dagger}_{i+1} \hat d^{\dagger}_{i+1}+
\hat e_i \hat d^{\dagger}_i \hat d_{i+1} \hat e_{i+1} + 
\hat d_i \hat e^{\dagger}_i \hat e^{\dagger}_{i+1} \hat d^{\dagger}_{i+1}
\nonumber\\
&+&
\hat d_i \hat e^{\dagger}_i \hat d_{i+1} \hat e_{i+1}) ] \}. 
\label{EQU19}
\end{eqnarray}
\begin{eqnarray}
\hat P_{10,i} &=& \frac{1}{4} \{ (\frac{1}{XV}+\frac{1}{UY}) [ (\frac{1}{XW}+
\frac{1}{UZ})(\hat a_i \hat b^{\dagger}_i\hat a_{i+1} \hat e^{\dagger}_{i+1}+ 
\hat a_i \hat b^{\dagger}_i \hat e_{i+1} \hat a^{\dagger}_{i+1} +
\hat b_i \hat a^{\dagger}_i \hat a_{i+1} \hat e^{\dagger}_{i+1} +
\hat b_i \hat a^{\dagger}_i \hat e_{i+1} \hat a^{\dagger}_{i+1}
\nonumber\\
&+&
\hat a_i \hat e^{\dagger}_i\hat a_{i+1} \hat b^{\dagger}_{i+1}+
\hat a_i \hat e^{\dagger}_i \hat b_{i+1} \hat a^{\dagger}_{i+1} + 
\hat e_i \hat a^{\dagger}_i \hat a_{i+1} \hat b^{\dagger}_{i+1} +
\hat e_i \hat a^{\dagger}_i \hat b_{i+1} \hat a^{\dagger}_{i+1}) +
(\frac{1}{XR}+\frac{1}{UQ})(\hat a_i \hat b^{\dagger}_i\hat a_{i+1}
\hat d^{\dagger}_{i+1}
\nonumber\\
&+&
\hat a_i \hat b^{\dagger}_i \hat d_{i+1} \hat a^{\dagger}_{i+1} +
\hat b_i \hat a^{\dagger}_i \hat a_{i+1} \hat d^{\dagger}_{i+1} +
\hat b_i \hat a^{\dagger}_i \hat d_{i+1} \hat a^{\dagger}_{i+1} +
\hat a_i \hat d^{\dagger}_i\hat a_{i+1} \hat b^{\dagger}_{i+1}+
\hat a_i \hat d^{\dagger}_i \hat b_{i+1} \hat a^{\dagger}_{i+1} + 
\hat d_i \hat a^{\dagger}_i \hat a_{i+1} \hat b^{\dagger}_{i+1}
\nonumber\\%
&+&
\hat d_i \hat a^{\dagger}_i \hat b_{i+1} \hat a^{\dagger}_{i+1}) +
(\frac{1}{YW}+\frac{1}{ZV})(\hat a_i \hat b^{\dagger}_i\hat b_{i+1}
\hat e^{\dagger}_{i+1}+ 
\hat a_i \hat b^{\dagger}_i \hat e_{i+1} \hat b^{\dagger}_{i+1} +
\hat b_i \hat a^{\dagger}_i \hat b_{i+1} \hat e^{\dagger}_{i+1} +
\hat b_i \hat a^{\dagger}_i \hat e_{i+1} \hat b^{\dagger}_{i+1}
\nonumber\\
&+&
\hat b_i \hat e^{\dagger}_i\hat a_{i+1} \hat b^{\dagger}_{i+1}+
\hat b_i \hat e^{\dagger}_i \hat b_{i+1} \hat a^{\dagger}_{i+1} + 
\hat e_i \hat b^{\dagger}_i \hat a_{i+1} \hat b^{\dagger}_{i+1} +
\hat e_i \hat b^{\dagger}_i \hat b_{i+1} \hat a^{\dagger}_{i+1}) +
(\frac{1}{YR}+\frac{1}{VQ})(\hat a_i \hat b^{\dagger}_i\hat b_{i+1}
\hat d^{\dagger}_{i+1}
\nonumber\\
&+&
\hat a_i \hat b^{\dagger}_i \hat d_{i+1} \hat b^{\dagger}_{i+1} +
\hat b_i \hat a^{\dagger}_i \hat b_{i+1} \hat d^{\dagger}_{i+1} +
\hat b_i \hat a^{\dagger}_i \hat d_{i+1} \hat b^{\dagger}_{i+1} +
\hat b_i \hat d^{\dagger}_i\hat a_{i+1} \hat b^{\dagger}_{i+1}+
\hat b_i \hat d^{\dagger}_i \hat b_{i+1} \hat a^{\dagger}_{i+1} + 
\hat d_i \hat b^{\dagger}_i \hat a_{i+1} \hat b^{\dagger}_{i+1}
\nonumber\\
&+&
\hat d_i \hat b^{\dagger}_i \hat b_{i+1} \hat a^{\dagger}_{i+1}) +
(\frac{1}{ZR}+\frac{1}{WQ})
(\hat a_i \hat b^{\dagger}_i\hat e_{i+1} \hat d^{\dagger}_{i+1}+ 
\hat a_i \hat b^{\dagger}_i \hat d_{i+1} \hat e^{\dagger}_{i+1} +
\hat b_i \hat a^{\dagger}_i \hat e_{i+1} \hat d^{\dagger}_{i+1} +
\hat b_i \hat a^{\dagger}_i \hat d_{i+1} \hat e^{\dagger}_{i+1}
\nonumber\\
&+&
\hat e_i \hat d^{\dagger}_i\hat a_{i+1} \hat b^{\dagger}_{i+1}+
\hat e_i \hat d^{\dagger}_i \hat b_{i+1} \hat a^{\dagger}_{i+1} + 
\hat d_i \hat e^{\dagger}_i \hat a_{i+1} \hat b^{\dagger}_{i+1} +
\hat d_i \hat e^{\dagger}_i \hat b_{i+1} \hat a^{\dagger}_{i+1})]+
(\frac{1}{XW}+\frac{1}{UZ}) [ (\frac{1}{XR}+\frac{1}{UQ})
\nonumber\\
&\times& 
(\hat a_i \hat e^{\dagger}_i\hat a_{i+1} \hat d^{\dagger}_{i+1} +
\hat a_i \hat e^{\dagger}_i \hat d_{i+1} \hat a^{\dagger}_{i+1} +
\hat e_i \hat a^{\dagger}_i \hat a_{i+1} \hat d^{\dagger}_{i+1} +
\hat e_i \hat a^{\dagger}_i \hat d_{i+1} \hat a^{\dagger}_{i+1} +
\hat a_i \hat d^{\dagger}_i\hat a_{i+1} \hat e^{\dagger}_{i+1} +
\hat a_i \hat d^{\dagger}_i \hat e_{i+1} \hat a^{\dagger}_{i+1}
\nonumber\\
&+& 
\hat d_i \hat a^{\dagger}_i \hat a_{i+1} \hat e^{\dagger}_{i+1} +
\hat d_i \hat a^{\dagger}_i \hat e_{i+1} \hat a^{\dagger}_{i+1}) +
(\frac{1}{YW}+\frac{1}{ZV})(\hat a_i \hat e^{\dagger}_i\hat b_{i+1}
\hat e^{\dagger}_{i+1}+ 
\hat a_i \hat e^{\dagger}_i \hat e_{i+1} \hat b^{\dagger}_{i+1} +
\hat e_i \hat a^{\dagger}_i \hat b_{i+1} \hat e^{\dagger}_{i+1}
\nonumber\\
&+&
\hat e_i \hat a^{\dagger}_i \hat e_{i+1} \hat b^{\dagger}_{i+1} +
\hat b_i \hat e^{\dagger}_i\hat a_{i+1} \hat e^{\dagger}_{i+1}+
\hat b_i \hat e^{\dagger}_i \hat e_{i+1} \hat a^{\dagger}_{i+1} + 
\hat e_i \hat b^{\dagger}_i \hat a_{i+1} \hat e^{\dagger}_{i+1} +
\hat e_i \hat b^{\dagger}_i \hat e_{i+1} \hat a^{\dagger}_{i+1}) +
(\frac{1}{YR}+\frac{1}{QV})
\nonumber\\
&\times& 
(\hat a_i \hat e^{\dagger}_i\hat b_{i+1} \hat d^{\dagger}_{i+1} +
\hat a_i \hat e^{\dagger}_i \hat d_{i+1} \hat b^{\dagger}_{i+1} +
\hat e_i \hat a^{\dagger}_i \hat b_{i+1} \hat d^{\dagger}_{i+1} +
\hat e_i \hat a^{\dagger}_i \hat d_{i+1} \hat b^{\dagger}_{i+1} +
\hat b_i \hat d^{\dagger}_i\hat a_{i+1} \hat e^{\dagger}_{i+1} +
\hat b_i \hat d^{\dagger}_i \hat e_{i+1} \hat a^{\dagger}_{i+1}
\nonumber\\
&+&
\hat d_i \hat b^{\dagger}_i \hat a_{i+1} \hat e^{\dagger}_{i+1} +
\hat d_i \hat b^{\dagger}_i \hat e_{i+1} \hat a^{\dagger}_{i+1}) +
(\frac{1}{ZR}+\frac{1}{QW})
(\hat a_i \hat e^{\dagger}_i\hat e_{i+1}
\hat d^{\dagger}_{i+1}+ 
\hat a_i \hat e^{\dagger}_i \hat d_{i+1} \hat e^{\dagger}_{i+1} +
\hat e_i \hat a^{\dagger}_i \hat e_{i+1} \hat d^{\dagger}_{i+1}
\nonumber\\
&+&
\hat e_i \hat a^{\dagger}_i \hat d_{i+1} \hat e^{\dagger}_{i+1} +
\hat e_i \hat d^{\dagger}_i\hat a_{i+1} \hat e^{\dagger}_{i+1}+
\hat e_i \hat d^{\dagger}_i \hat e_{i+1} \hat a^{\dagger}_{i+1} + 
\hat d_i \hat e^{\dagger}_i \hat a_{i+1} \hat e^{\dagger}_{i+1} +
\hat d_i \hat e^{\dagger}_i \hat e_{i+1} \hat a^{\dagger}_{i+1})] +
(\frac{1}{XR}+\frac{1}{QU})
\nonumber\\
&\times& [ (\frac{1}{YW}+\frac{1}{ZV})(
\hat a_i \hat d^{\dagger}_i\hat b_{i+1} \hat e^{\dagger}_{i+1} +
\hat a_i \hat d^{\dagger}_i \hat e_{i+1} \hat b^{\dagger}_{i+1} +
\hat d_i \hat a^{\dagger}_i \hat b_{i+1} \hat e^{\dagger}_{i+1} +
\hat d_i \hat a^{\dagger}_i \hat e_{i+1} \hat b^{\dagger}_{i+1} +
\hat b_i \hat e^{\dagger}_i\hat a_{i+1} \hat d^{\dagger}_{i+1}
\nonumber\\
&+&
\hat b_i \hat e^{\dagger}_i \hat d_{i+1} \hat a^{\dagger}_{i+1} +
\hat e_i \hat b^{\dagger}_i \hat a_{i+1} \hat d^{\dagger}_{i+1} +
\hat e_i \hat b^{\dagger}_i \hat d_{i+1} \hat a^{\dagger}_{i+1}) +
(\frac{1}{YR}+\frac{1}{QV})(
\hat a_i \hat d^{\dagger}_i\hat b_{i+1} \hat d^{\dagger}_{i+1} +
\hat a_i \hat d^{\dagger}_i \hat d_{i+1} \hat b^{\dagger}_{i+1}
\nonumber\\
&+&
\hat d_i \hat a^{\dagger}_i \hat b_{i+1} \hat d^{\dagger}_{i+1} +
\hat d_i \hat a^{\dagger}_i \hat d_{i+1} \hat b^{\dagger}_{i+1} +
\hat b_i \hat d^{\dagger}_i\hat a_{i+1} \hat d^{\dagger}_{i+1} +
\hat b_i \hat d^{\dagger}_i\hat d_{i+1} \hat a^{\dagger}_{i+1} +
\hat d_i \hat b^{\dagger}_i\hat a_{i+1} \hat d^{\dagger}_{i+1} +
\hat d_i \hat b^{\dagger}_i\hat d_{i+1} \hat a^{\dagger}_{i+1})
\nonumber\\
&+&(\frac{1}{ZR}+\frac{1}{QW})(
\hat a_i \hat d^{\dagger}_i \hat e_{i+1} \hat d^{\dagger}_{i+1} +
\hat a_i \hat d^{\dagger}_i \hat d_{i+1} \hat e^{\dagger}_{i+1} +
\hat d_i \hat a^{\dagger}_i \hat e_{i+1} \hat d^{\dagger}_{i+1} +
\hat d_i \hat a^{\dagger}_i \hat d_{i+1} \hat e^{\dagger}_{i+1} +
\hat e_i \hat d^{\dagger}_i\hat a_{i+1} \hat d^{\dagger}_{i+1}
\nonumber\\
&+&
\hat e_i \hat d^{\dagger}_i\hat d_{i+1} \hat a^{\dagger}_{i+1} +
\hat d_i \hat e^{\dagger}_i\hat a_{i+1} \hat d^{\dagger}_{i+1} +
\hat d_i \hat e^{\dagger}_i\hat d_{i+1} \hat a^{\dagger}_{i+1})] +
(\frac{1}{YW}+\frac{1}{VZ}) [ (\frac{1}{YR}+\frac{1}{VQ})(
\hat b_i \hat e^{\dagger}_i \hat b_{i+1} \hat d^{\dagger}_{i+1}
\nonumber\\
&+&
\hat b_i \hat e^{\dagger}_i \hat d_{i+1} \hat b^{\dagger}_{i+1} +
\hat e_i \hat b^{\dagger}_i \hat b_{i+1} \hat d^{\dagger}_{i+1} +
\hat e_i \hat b^{\dagger}_i \hat d_{i+1} \hat b^{\dagger}_{i+1} +
\hat b_i \hat d^{\dagger}_i\hat b_{i+1} \hat e^{\dagger}_{i+1} +
\hat b_i \hat d^{\dagger}_i\hat e_{i+1} \hat b^{\dagger}_{i+1} +
\hat d_i \hat b^{\dagger}_i\hat b_{i+1} \hat e^{\dagger}_{i+1}
\nonumber\\
&+& 
\hat d_i \hat b^{\dagger}_i\hat e_{i+1} \hat b^{\dagger}_{i+1}) +
(\frac{1}{ZR}+\frac{1}{WQ})(
\hat b_i \hat e^{\dagger}_i \hat e_{i+1} \hat d^{\dagger}_{i+1} +
\hat b_i \hat e^{\dagger}_i \hat d_{i+1} \hat e^{\dagger}_{i+1} +
\hat e_i \hat b^{\dagger}_i \hat e_{i+1} \hat d^{\dagger}_{i+1} +
\hat e_i \hat b^{\dagger}_i \hat d_{i+1} \hat e^{\dagger}_{i+1}
\nonumber\\
&+&
\hat e_i \hat d^{\dagger}_i\hat b_{i+1} \hat e^{\dagger}_{i+1} +
\hat e_i \hat d^{\dagger}_i\hat e_{i+1} \hat b^{\dagger}_{i+1} +
\hat d_i \hat e^{\dagger}_i\hat b_{i+1} \hat e^{\dagger}_{i+1} +
\hat d_i \hat e^{\dagger}_i\hat e_{i+1} \hat b^{\dagger}_{i+1})] +
(\frac{1}{YR}+\frac{1}{VQ}) [ (\frac{1}{ZR}+\frac{1}{WQ})
\nonumber\\
&\times&(
\hat b_i \hat d^{\dagger}_i \hat e_{i+1} \hat d^{\dagger}_{i+1} +
\hat b_i \hat d^{\dagger}_i \hat d_{i+1} \hat e^{\dagger}_{i+1} +
\hat d_i \hat b^{\dagger}_i \hat e_{i+1} \hat d^{\dagger}_{i+1} +
\hat d_i \hat b^{\dagger}_i \hat d_{i+1} \hat e^{\dagger}_{i+1} +
\hat e_i \hat d^{\dagger}_i\hat b_{i+1} \hat d^{\dagger}_{i+1} +
\hat e_i \hat d^{\dagger}_i\hat d_{i+1} \hat b^{\dagger}_{i+1}
\nonumber\\
&+&
\hat d_i \hat e^{\dagger}_i\hat b_{i+1} \hat d^{\dagger}_{i+1} +
\hat d_i \hat e^{\dagger}_i\hat d_{i+1} \hat b^{\dagger}_{i+1})] \}
\label{EQU20}
\end{eqnarray}



\begin{references}

\bibitem{JW1}
Z. Gulacsi, Jordan-Wigner transformation constructed for spinful fermions at
S=1/2 spins in one dimension, Phil. Mag. 105, 71 (2025), Available at
https://doi.org/10.1080/14786435.2024.2392703

\bibitem{JW2}
Z. Gulacsi, Jordan–Wigner transformation constructed for spinful fermions at
spin-1/2 in two dimensions, Accepted for publication at Phil. Mag. (2025),
Available online at
https://doi.org/10.1080/14786435.2025.2465731

\bibitem{JW3}
P. Jordan an E. Wigner, \"Uber das Paulische \"Aquivalenzverbot,
Z. Physik 47, 631 (1928), Available at:
https://doi.org/10.1007/BF01331938

\bibitem{JW4}
E. Lieb, T. Schultz, and D. Mattis, Two soluble models of an antiferromagnetic
chain, Annals of Physics 16, 407 (1961),\\
Available at
https://www.sciencedirect.com/science/article/abs/pii/0003491661901154

\bibitem{JW5}
M. Azzouz, Interchain-coupling effect on the one dimensional spin-1/2
antiferromagnetic Heisenberg model, Phys. Rev. B48, 6136 (1993),\\
Available at
https://journals.aps.org/prb/pdf/10.1103/PhysRevB.48.6136

\bibitem{JW6}
Y. R. Wang, Ground state of the two-dimensional antiferromagnetic Heisenberg
model studied using an extended Wigner-Jordan transformation,
Phys. Rev. B43, 3786 (1991)\\
Available at
https://journals.aps.org/prb/pdf/10.1103/PhysRevB.43.3786


\bibitem{JW7}
E. Fradkin, Jordan-Wigner transformation for quantum-spin systems in two
dimensions and fractional statistics, Phys. Rev. Lett. 63, 322 (1989)\\
Available at
https://journals.aps.org/prl/pdf/10.1103/PhysRevLett.63.322

\bibitem{JW8}
C. B. Batista and G. Ortiz, Generalized Jordan-Wigner transformation,
Phys. Rev. Lett. 86, 1082 (2001),\\
Available at
https://journals.aps.org/prl/abstract/10.1103/PhysRevLett.86.1082


\bibitem{JW30}
N. Kucska and Z. Gulacsi, Spin-orbit interactions may relax the rigid condi-
tions leading to flat bands, Phys. Rev. B105, 085103 (2022),
Available at https://doi.org/10.1103/PhysRevB.105.085103

\bibitem{JW300}
T. Tsuchimochi, Spinflip configuration interaction singles with exact spin-
projection: Theory and applications to strongly correlated systems,
J. Chem. Phys. 143, 144114 (2015), Available at
https://doi.org/10.1063/1.4933113
  
\bibitem{JW3a}
X. Yang, S. Biswas, S. Lu, M. Randeria and Y. M. Lu, Pairing symmetry and
fermion projective symmetry groups, SciPost Phys. 17, 161 (2024), Available at
https://doi.org/10.21468/SciPostPhys.17.6.161

\bibitem{JW3aa}
M. Bercx, J. S. Hofmann, F. F. Assaad and T. C. Lang,  Spontaneous
particle-hole symmetry breaking of correlated fermions on the Lieb lattice,
Phys. Rev. B 95, 035108 (2017), Available at
 https://doi.org/10.1103/PhysRevB.95.035108

\bibitem{JW3b}
S. Imai and N. Tsuji, Quantum many-body scars with unconventional
superconducting pairing symmetries via multibody interactions, Phys. Rev.
Research 7, 013064 (2025), Available at
https://doi.org/10.1103/PhysRevResearch.7.013064
  
\bibitem{JW3c}
S. D. Lundemo and A. Sudbo, Topological superconductivity induced by a Kitaev spin liquid, Phys. Rev. B109, 184508 (2024), Available at 
https://doi.org/10.1103/PhysRevB.109.184508

\bibitem{JW3d}
K. Kang, B. Shen, Y. Qiu, Y. Zeng, Z. Xia, K. Watanabe, T. Taniguchi,
J. Shan and K. F. Mak, Evidence of the fractional quantum spin Hall
effect in moiré MoTe2, Nature 628, 522 (2024), Available at
https://doi.org/10.1038/s41586-024-07214-5

\bibitem{JW3e}
Y. Xie, A. T. Pierce, J. M. Park, D. E. Parker, E. Khalaf, P. Ledwith,
Y. Cao, S. H. Lee, S. Chen, P. R. Forrester, K. Watanabe, T. Taniguchi,
A. Vishwanath, P. Jarillo-Herrero and  A. Yacoby, Fractional Chern insulators
in magic-angle twisted bilayer graphene, Nature 600, 439 (2021), Available at
https://doi.org/10.1038/s41586-021-04002-3

\bibitem{JW3f}
El-Nabulsi, R.A. On, Generalized Fractional Spin, Fractional Angular
Momentum, Fractional Momentum Operators in Quantum Mechanics,
Few-Body Syst. 61, 25 (2020), Available at
https://doi.org/10.1007/s00601-020-01558-0

\bibitem{JW3g}
A. Hokkyo, Rigorous Test for Quantum Integrability and Nonintegrability, 2025
arXiv:2501.18400, Available at  	
https://doi.org/10.48550/arXiv.2501.18400

\bibitem{JW3h}
Z. Gulacsi and D. Vollhardt, Exact Insulating and Conducting Ground States
of a Periodic Anderson Model in Three Dimensions, Phys. Rev. Lett. 91,
186401, (2003), Available at https://doi.org/10.1103/PhysRevLett.91.186401

\bibitem{JW9}
M. Kollar and D. Vollhardt, Correlated hopping of electrons: Eﬀect on the
Brinkman-Rice transition  and the stability of metallic ferromagnetism,
Phys. Rev. B 63, 045107 (2001), Available at
https://doi.org/10.1103/PhysRevB.63.045107


\bibitem{JW10}
A. Lau, J. van den Brink and C. Ortix, Topological mirror insulators
in one dimension, Phys. Rev B 94, 165164 (2016), Available at
https://doi.org/10.1103/PhysRevB.94.165164

\bibitem{JW11}
J. E. Birkholz and V. Meden, Spin-orbit coupling effects in one-dimensional
ballistic quantum wires, J. Phys.: Condensed Matter 20, 085226 (2008),
Available at https://doi.org/10.1088/0953-8984/20/8/085226

\bibitem{JW12}
Z. G. Yu, Spin-orbit coupling and its effects in organic solids,
Phys. Rev. B 85, 115201 (2012), Available at
https://doi.org/10.1103/PhysRevB.85.115201

  

\bibitem{JW13}
N. Kaushal, J. Herbrych, A. Nocera, G. Alvarez, A. Moreo, F. A. Reboredo,
and  E. Dagotto, Density matrix renormalization group study of a
three-orbital Hubbard model with spin-orbit coupling in one dimension,
Phys. Rev. B 96, 155111 (2017), Available at
https://doi.org/10.1103/PhysRevB.96.155111

\bibitem{JW14}
Y. Chen, Yi-H. Tian, R.-Q. He and Z.-Yi Lu, Spin-orbit coupling effects
on orbital-selective correlations in a three-orbital model, arXiv:2503.14435
Available at https://doi.org/10.48550/arXiv.2503.14435

\bibitem{JW15}
N. Kucska and Z. Gulacsi, Spin-orbit interactions may relax the rigid
conditions leading to flat bands, Phys. Rev. B105, 085103 (2022),
Available at
https://doi.org/10.1103/PhysRevB.105.085103

\bibitem{JW16}
S. Mohapatra and A. Singh, Spin waves and stability of zigzag order
in the Hubbard model with spin-dependent hopping terms: Application
to the honeycomb lattice compounds Na2IrO3 and a-RuCl3,
Jour. Magn. Magn. Matter 479 (2019) 229, Available at
https://doi.org/10.1016/j.jmmm.2019.02.013


\bibitem{JW17}
K. Kuboki, Phase Separation Induced by Density-Dependent Hopping Terms,
J. Phys. Soc. Jpn. 93, 035001 (2024), Available at
https://doi.org/10.7566/JPSJ.93.035001

\bibitem{JW18}
T. Westerhout and M. I. Katsnelson, The role of correlated hopping in
many-body physics of flat-band systems: Nagaoka ferromagnetism,
Phys. Rev. B 106, L041104 (2022), Available at
https://doi.org/10.1103/PhysRevB.106.L041104

\bibitem{JW19}
W. Chen, J. Zhang and H. Ding, Ground-state instabilities in a Hubbard-type
chain with particular correlated hopping at non-half-filling,
Results in Physics 49 (2023) 106472, Available at
https://doi.org/10.1016/j.rinp.2023.106472

\bibitem{JW20}
P. Dhuria, S. S. Bhamra and J. S. Hundal, The study of Correlated Barrier
Hopping (CBH) conduction mechanism and modulus spectroscopy of
YFe0.5Co0.5O3 compounds, Physica B 677 (2024) 415696, Available at
https://10.1016/j.physb.2024.415696



\bibitem{JW21}
X. Sun, W. Chen and H. Ding, Phase diagram of the Hubbard chain with
symmetric density-dependent hopping, Results in Physics 65 (2024) 107983,
Available at https://doi.org/10.1016/j.rinp.2024.107983

\bibitem{JW22}
P. Roura-Bas and A. A. Aligia, Phase diagram of the ionic Hubbard model
with density-dependent hopping, Phys. Rev. B 108, 115132 (2023),
Available at https://doi.org/10.1103/PhysRevB.108.115132

\bibitem{JW23}
V. Lienhard, P. Scholl, S. Weber, D. Barredo, S. de Léséleuc, R. Bai,
N. Lang, M. Fleischhauer, H. P. Büchler, T. Lahaye and A. Browaeys,
Realization of a Density-Dependent Peierls Phase in a Synthetic,
Spin-Orbit Coupled Rydberg System, Phys. Rev. X 10, 021031 (2020),
Available at https://doi.org/10.1103/PhysRevX.10.021031

\bibitem{JW24}
P. Xu, T.-S. Deng, W. Zheng and H. Zhai, Density-dependent spin-orbit
coupling in degenerate quantum gases, Phys. Rev. A 103, L061302 (2021),
Available at https://doi.org/10.1103/PhysRevA.103.L061302

\bibitem{JW25}
S. Phuntsho, Doping-Driven Modulation of Spin–Orbit Coupling, Spin
Textures, and Rashba–Edelstein Response in Chiral Tellurium: A
First-Principles Study, arXiv:2503.01685, Available at:
https://doi.org/10.48550/arXiv.2503.01685

\bibitem{JW26}
N. Bovenzi, S. Caprara, M. Grilli, R. Raimondi, N. Scopigno and G. Seibold,
Density inhomogeneities and Rashba spin-orbit coupling interplay in
oxide interfaces, Jour. of Phys. and Chem. of Solids 128 (2019) 118,
Available at https://dx.doi.org/10.1016/j.jpcs.2017.09.013


\bibitem{JW27}
K. Jiang, Correlation Renormalized and Induced Spin-Orbit Coupling,
Chinese Physics Letters 40, 017102 (2023), Available at
https://doi.org/10.1088/0256-307X/40/1/017102

\bibitem{JW28}
S. Kim, K. Ueda, G. Go, P.-H. Jang, K.-J. Lee, et al., Correlation of
the Dzyaloshinskii–Moriya interaction with Heisenberg exchange and orbital
asphericity, Nature Comm. 9, 1648 (2018), Available at
https://doi.org/10.1038/s41467-018-04017-x





\end{references}
\end{document}